\documentclass[prd,aps,floats,amssymb,preprint,nofootinbib]{revtex4}

\usepackage{epsfig}
\usepackage{amssymb}
\usepackage{bbm}
\usepackage{color}
\usepackage{ulem}

\usepackage{enumitem}
\usepackage{graphicx}
\usepackage{subfigure}
\usepackage{amsmath}
\usepackage{bbm}
\usepackage{color}

\newcommand{\be}{\begin{equation}}
\newcommand{\ee}{\end{equation}}
\newcommand{\bea}{\begin{eqnarray}}
\newcommand{\eea}{\end{eqnarray}}
\newcommand{\beas}{\begin{eqnarray*}}
\newcommand{\eeas}{\end{eqnarray*}}
\newcommand{\nn}{\nonumber}

\newcommand{\half}{{1\over 2}}

\newcommand\di{\partial}

\newcommand{\dd}{\text{d}}

\begin{document}


\title{Multi-field galileons and higher co-dimension branes}
\author{Kurt Hinterbichler\footnote{kurthi@physics.upenn.edu}, Mark Trodden\footnote{trodden@physics.upenn.edu} and Daniel Wesley\footnote{dwes@sas.upenn.edu}}

\affiliation{Center for Particle Cosmology, Department of Physics and Astronomy, University of Pennsylvania,
Philadelphia, Pennsylvania 19104, USA}

\date{\today}

\begin{abstract}
In the decoupling limit, the DGP model reduces to the theory of a scalar field $\pi$, with interactions including a
specific cubic self-interaction - the galileon term.  This term, and its quartic and quintic generalizations, can be thought of as arising from a probe $3$-brane in a $5$-dimensional bulk with Lovelock terms on the brane and in the bulk.  We study multi-field generalizations of the galileon, and extend this probe brane view to higher co-dimensions.  We derive an extremely restrictive theory of multiple galileon fields, interacting through a quartic term controlled by a single coupling, and trace its origin to the induced brane terms coming from Lovelock invariants in the higher co-dimension bulk.   We explore some properties of this theory, finding de Sitter like self accelerating solutions.  These solutions  have ghosts if and only if the flat space theory does not have ghosts.  Finally, we prove a general non-renormalization theorem: multi-field galileons are not renormalized quantum mechanically to any loop in perturbation theory. 

\end{abstract}

\maketitle

\setcounter{footnote}{0}

\section{Introduction}
A particularly fruitful way of extending both the standard models of particle physics and cosmology is the hypothesis of extra spatial dimensions beyond the three that manifest themselves in everyday physics. Historically, such ideas have provided a tantalizing possibility of unifying the
basic forces through the geometry and topology of the extra-dimensional manifold, and in recent years, have been the basis for attempts
to tackle the hierarchy problem. In this latter incarnation, a crucial insight has been the realization that different forces may operate in 
different dimensionalities, by confining the standard model particles to a $3+1$-dimensional submanifold - the brane - while gravity probes
the entire spacetime - the bulk - due to the equivalence principle. Such constructions allow, among other unusual features, for infinite extra
dimensions, in contrast to the more usual compactified theories.

In the case of a single extra dimension, a further refinement was introduced in~\cite{Dvali:2000hr}, where
a separate induced gravity term was introduced on the brane. The resulting $4+1$-dimensional action
\be
S=\frac{M_5^3}{2}\int d^5x\, \sqrt{-G}\,\  R[G] + \frac{M_4^2}{2}\int d^4x\, \sqrt{-g}\, \ R[g]
\label{DGPaction}
\ee
is known as the DGP (Dvali-Gabadadze-Porrati) model, and yields a rich and dramatic phenomenology, with, for example, a branch of $4$-dimensional cosmological solutions which self-accelerate
at late times, and a set of predictions for upcoming missions which will perform local tests of gravity.

It is possible to derive a 4-d effective action for the DGP model by integrating out the bulk.  It has been claimed \cite{Luty:2003vm,Nicolis:2004qq} that a decoupling limit for DGP exists, in which the 4-d effective action reduces to a theory of a single scalar $\pi$, representing the position of the brane in the extra dimension, with a cubic self-interaction term $\sim (\partial\pi)^2\square \pi$ (though this claim is not without controversy, see for example \cite{Gabadadze:2006tf}).  This term has the properties that its field equations are second order (despite the fact that the lagrangian is higher order), which is important for avoiding ghosts.   It is also invariant (up to a total derivative) under the following {\it galilean} transformation,
\be \label{galileoninvarianceold}
\pi (x) \rightarrow \pi(x) + c + b_\mu x^\mu \ ,
\ee
with $c$ and $b_{\mu}$ constants.  

These properties are interesting in their own right, and terms that generalize the cubic DGP term studied (without considering a possible higher dimensional origin) in~\cite{Nicolis:2008in} are referred to as \textit{galileons}.  Requiring the invariance (\ref{galileoninvarianceold}) forces the equations of motion to contain at least two derivatives acting on each field, and there exists a 
set of terms that lead to such a form with exactly two derivatives on each field (in fact, the absence of ghosts in a non-linear regime demands that there be at most two derivatives on each field).  These are the terms that were classified in \cite{Nicolis:2008in}, and take the schematic form
\be \label{Ln}
{\cal L}_n \sim \partial \pi \, \partial \pi\, (\di^2 \pi)^{n-2},
\ee
with suitable Lorentz contractions and dimensionful coefficients.  In $d$ spacetime dimensions there are $d$ such terms, corresponding to $n= 2, \dots, d+1$.  The $n=2$ term is just the usual kinetic term $(\partial \pi)^2$, the $n=3$ case is the DGP term $(\di \pi)^2 \Box \pi$, and the higher terms generalize these.  

These terms have appeared in various contexts apart from DGP; for example the $n=4,5$ terms seem to appear in the decoupling limit of an interesting interacting theory of Lorentz invariant massive gravity~\cite{deRham:2010gu}.  They have been generalized to curved space~\cite{Deffayet:2009wt,Deffayet:2009mn}, identified as possible ghost-free modifications of gravity and cosmology~\cite{Nicolis:2008in,Silva:2009km,Kobayashi:2010wa,DeFelice:2010pv,Chow:2009fm,Agarwal:2009gy}, and used to build alternatives to inflation \cite{Creminelli:2010ba} and dark energy \cite{Gannouji:2010au,Deffayet:2010qz}.  

Another remarkable fact, which we will prove for a more general multi-field model in section~\ref{nonrenormalizesection}, is that the ${\cal L}_n$
terms above do not get renormalized upon loop corrections, so that their classical values can be trusted quantum-mechanically.  Also, from an effective field theory point of view, there can exist regimes in which only these galileon terms are important.

It is natural to consider whether the successes of the DGP model can be extended and improved in models in which the bulk has higher
co-dimension, and whether the drawbacks of the $5$-dimensional approach, such as the ghost problem in the accelerating branch,
might be ameliorated in such a setting. Since our understanding of the complexities of the DGP model has arisen primarily through
the development of a $4$-dimensional effective theory in a decoupling limit, one might hope to achieve a similar understanding of theories with larger co-dimension.  This is the aim of this paper.

We do not consider the full higher co-dimension DGP or a decoupling limit thereof.  Instead, we are interested in generalizing the galileon actions to multiple fields and exploring the probe brane-world view of these terms, extending the work of~\cite{deRham:2010eu} on the single field case.  The theory which emerges from the brane construction in co-dimension $N$ has an internal $SO(N)$ symmetry in addition to the galilean symmetry.  This is extremely restrictive, and in four dimensions it turns out that there is a single non-linear term compatible with it.  This makes for a fascinating four dimensional field description; a scalar field theory with a single allowed coupling, which receives no quantum corrections.

\section{Single-field galileons and generalizations}

In co-dimension one, the decoupling limit of DGP consists of a $4$-dimensional effective theory of gravity coupled to a single scalar field $\pi$, representing
the bending mode of the brane in the fifth dimension. The $\pi$ field self-interaction includes a cubic self-interaction $\sim (\partial\pi)^2\square\pi$, which has the following two properties:
\begin{enumerate} 
\item The field equations are second order, 
\item The terms are invariant up to a total derivative under the internal galilean transformations $\pi\rightarrow \pi+c+b_\mu x^\mu$, where $c,b_\mu$ are arbitrary real constants.
\end{enumerate}

In \cite{Nicolis:2008in}, this was generalized, and all possible lagrangian terms for a single scalar with these two properties were classified in all dimensions.  They are called galileon terms, and there exists a single galileon lagrangian at each order in $\pi$, where ``order" refers to the number of copies of $\pi$ that appear in the term.   
For $n\geq 1$, the $(n+1)$-th order galileon lagrangian is
\be
\label{galileon2} 
{\cal L}_{n+1}=n\eta^{\mu_1\nu_1\mu_2\nu_2\cdots\mu_n\nu_n}\left( \partial_{\mu_1}\pi\partial_{\nu_1}\pi\partial_{\mu_2}\partial_{\nu_2}\pi\cdots\partial_{\mu_n}\partial_{\nu_n}\pi\right),
\ee 
where 
\be
\label{tensor} 
\eta^{\mu_1\nu_1\mu_2\nu_2\cdots\mu_n\nu_n}\equiv{1\over n!}\sum_p\left(-1\right)^{p}\eta^{\mu_1p(\nu_1)}\eta^{\mu_2p(\nu_2)}\cdots\eta^{\mu_np(\nu_n)} \ .
\ee 
The sum in~(\ref{tensor}) is over all permutations of the $\nu$ indices, with $(-1)^p$ the sign of the permutation.  The tensor~(\ref{tensor}) is anti-symmetric in the $\mu$ indices, anti-symmetric the $\nu$ indices, and symmetric under interchange of any $\mu,\nu$ pair with any other.  These lagrangians are unique up to total derivatives and overall constants.   Because of the anti-symmetry requirement on $\eta$, only the first $n$ of these galileons are non-trivial in $n$-dimensions.  In addition, the tadpole term, $\pi$, is galilean invariant, and we therefore include it as the first-order galileon.  

Thus, at the first few orders, we have 
\bea
 {\cal L}_1&=&\pi, \\ \nn
 {\cal L}_2&=&[\pi^2], \\ \nn
{\cal L}_3&=&[\pi^2][\Pi]-[\pi^3], \\ \nn
{\cal L}_4&=&\half[\pi^2][\Pi]^2-[\pi^3][\Pi]+[\pi^4]-\half[\pi^2][\Pi^2], \\ \nn
{\cal L}_5&=&{1\over 6}[\pi^2][\Pi]^3-{1\over 2}[\pi^3][\Pi]^2+[\pi^4][\Pi]-[\pi^5]+{1\over 3}[\pi^2][\Pi^3]-{1\over 2}[\pi^2][\Pi][\Pi^2]+{1\over 2}[\pi^3][\Pi^2]
\ .
\eea
We have used the notation $\Pi$ for the matrix of partials $\Pi_{\mu\nu}\equiv\partial_{\mu}\partial_\nu\pi$, and $[\Pi^n]\equiv Tr(\Pi^n)$, e.g. $[\Pi]=\square\pi$, $[\Pi^2]=\partial_\mu\partial_\nu\pi\partial^\mu\partial^\nu\pi$, and $[\pi^n]\equiv \partial\pi\cdot\Pi^{n-2}\cdot\partial\pi$, i.e. $[\pi^2]=\partial_\mu\pi\partial^\mu\pi$, $[\pi^3]=\partial_\mu\pi\partial^\mu\partial^\nu\pi\partial_\nu\pi$.  The above terms are the only ones which are non-vanishing in four dimensions.  The second is the standard kinetic term for a scalar, while the third is the DGP $\pi$-lagrangian (up to a total derivative).

The equations of motion derived from (\ref{galileon2}) are
\be 
\label{galileoneom} 
{\cal E}_{n+1}\equiv {\delta{\cal L}_{n+1}\over \delta \pi}=-n(n+1)\eta^{\mu_1\nu_1\mu_2\nu_2\cdots\mu_n\nu_n}\left( \partial_{\mu_1}\partial_{\nu_1}\pi\partial_{\mu_2}\partial_{\nu_2}\pi\cdots\partial_{\mu_n}\partial_{\nu_n}\pi\right)=0 \ ,
\ee
and are second order, as advertised\footnote{Beyond their second order nature, these lagrangians possess a number of other interesting properties.  Under the shift symmetry $\pi\rightarrow \pi+\epsilon$, the Noether current is
\be 
j^\mu_{n+1}=n(n+1)\eta^{\mu\nu_1\mu_2\nu_2\cdots\mu_n\nu_n}\left( \partial_{\nu_1}\pi\partial_{\mu_2}\partial_{\nu_2}\pi\cdots\partial_{\mu_n}\partial_{\nu_n}\pi\right) \ .
\ee
Shift symmetry implies that the equations of motion are equivalent to the conservation of this current,
\be 
{\cal E}_{n+1}=-\partial_\mu j^\mu_{n+1} \ .
\ee
However, the Noether current itself can also be written as a derivative 
\be 
j_{n+1}^\mu=\partial_\nu j^{\mu\nu}_{n+1} \ ,
\ee
where there are many possibilities for $j_{n+1}^\mu$, two examples of which are 
\bea 
j^{\mu\nu}_{n+1}&=&n(n+1)\eta^{\mu\nu\mu_2\nu_2\cdots\mu_n\nu_n}\left(\pi\partial_{\mu_2}\partial_{\nu_2}\pi\cdots\partial_{\mu_n}\partial_{\nu_n}\pi\right) \ , \\ 
j^{\mu\nu}_{n+1}&=&-n(n+1)\eta^{\mu\nu\mu_2\nu_2\cdots\mu_n\nu_n}\left( \partial_{\mu_2}\pi\partial_{\nu_2}\pi\partial_{\mu_3}\partial_{\nu_3}\pi\cdots\partial_{\mu_n}\partial_{\nu_n}\pi\right) \ .
\eea
Thus the equations of motion can in fact be written as a double total derivative,
\be  
{\cal E}_{n+1}=-\partial_\mu\partial_\nu j_{n+1}^{\mu\nu} \ .
\ee}.  

The first few orders of the equations of motion are 
\bea{\cal E}_1&=&1, \\
 {\cal E}_2&=&-2[\Pi], \\
{\cal E}_3&=& -3\left([\Pi]^2-[\Pi^2]\right),  \\
{\cal E}_4&=& -2\left([\Pi]^3+2[\Pi^3]-3[\Pi][\Pi^2]\right),  \\
{\cal E}_5&=& -{5\over 6}\left([\Pi]^4-6[\Pi^4]+8[\Pi][\Pi^3]-6[\Pi]^2[\Pi^2]+3[\Pi^2]^2\right) \ .
\eea

By adding a total derivative, and by using the following identity for the $\eta$ symbol in ${\cal L}_{n+1}$
\be  
\eta^{\mu_1\nu_1\ldots\mu_n\nu_n}={1\over n}\left(\eta^{\mu_1\nu_1}\eta^{\mu_2 \nu_2\ldots\mu_n \nu_n}-\eta^{\mu_1\nu_2}\eta^{\mu_2 \nu_1\mu_2\nu_3\ldots\mu_n\nu_n}+\cdots+(-1)^n\eta^{\mu_1\nu_n}\eta^{\mu_2\nu_1\ldots\mu_n\nu_{n-1}}\right) \ ,
\ee
the galileon lagrangians can be brought into a (sometimes more useful) different form, which illustrates that the $(n+1)$-th order lagrangian is just $(\partial\pi)^2$ times the $n$-th order equations of motion,
\be 
{\cal L}_{n+1}=-{n+1\over 2n(n-1)}(\partial\pi)^2{\cal E}_{n}-{n-1\over 2}\partial_{\mu_1}\left[(\partial\pi)^2\eta^{\mu_1\nu_1\cdots\mu_{n-1}\nu_{n-1}}\partial_{\nu_1}\pi\partial_{\mu_2}\partial_{\nu_2}\pi\cdots\partial_{\mu_{n-1}}\partial_{\nu_{n-1}}\pi\right] \ .
\label{simplifiedgalileon}
\ee
From the simplified form~(\ref{simplifiedgalileon}) we can see that ${\cal L}_3$, for example, takes the usual galileon form $(\partial\pi)^2\Box\pi$.

These galileon actions can be generalized to the multi-field case, where there is a multiplet $\pi^I$ of fields\footnote{As we put the finishing touches to this paper, several preprints appeared which also discuss generalizations to the galileons \cite{Deffayet:2010zh}, \cite{Padilla:2010ir}, \cite{Padilla:2010de}.}.  The action in this case can be written
\be \label{generaltermspre} {\cal L}_{n+1}= S_{I_1I_2\cdots I_{n+1}}\eta^{\mu_1\nu_1\mu_2\nu_2\cdots\mu_n\nu_n}\left(\pi^{I_{n+1}} \partial_{\mu_1}\partial_{\nu_1}\pi^{I_1}\partial_{\mu_2}\partial_{\nu_2}\pi^{I_2}\cdots\partial_{\mu_n}\partial_{\nu_n}\pi^{I_{n}}\right),
\ee
with $S_{I_1I_2\cdots I_{n+1}}$ a symmetric constant tensor.  This is invariant under under individual galilean transformations for each field, $\pi^I\rightarrow \pi^I+c^I+b^I_\mu x^\mu$, and the equations of motion are second order,
\be 
{\cal E}_{I}\equiv {\delta{\cal L}\over \delta \pi^I}=(n+1)S_{II_1I_2\cdots I_{n}}\eta^{\mu_1\nu_1\mu_2\nu_2\cdots\mu_n\nu_n}\left(\partial_{\mu_1}\partial_{\nu_1}\pi^{I_1}\partial_{\mu_2}\partial_{\nu_2}\pi^{I_2}\cdots\partial_{\mu_n}\partial_{\nu_n}\pi^{I_{n}}\right).
\ee

The theory containing these galilean-invariant operators is not renormalizable, i.e. it is an effective field theory with a cutoff $\Lambda$, above which some UV completion is required.   As was mentioned in the introduction, the ${\cal L}_n$ terms above do not get renormalized upon loop corrections, so that their classical values can be trusted quantum-mechanically (see section~\ref{nonrenormalizesection}).  The structure of the one-loop effective action (in $3+1$ dimensions) is, schematically\footnote{Strictly speaking, quantum effects calculable solely within the effective theory are only those associated with log-divergences.  Power-divergences are regularization dependent, and depend upon some UV completion or matching condition.  In dimensional regularization with minimal subtraction they do not even show up, corresponding to making a special and optimistic assumption about the UV completion, i.e. that power-law divergences are precisely cancelled somehow by the UV contributions.  However, it is important to stress that the conclusions about the galileon lagrangian are true even in the presence of generic power divergences, i.e. even with a generic UV completion.} \cite{Nicolis:2004qq},
\be\label{quantumaction}
\Gamma\sim \sum_m \left[ \Lambda^4 + \Lambda^2 \di^2 +
\di^4 \log \left(\frac{\di^2}{\Lambda^2}\right) \right] \left( \frac{\di \di \pi}{\Lambda^3} \right)^m  \ .
\ee

One should consider quantum effects within the effective theory, since there are other operators of the same dimension that might compete with the galileon terms.  However, there can exist interesting regimes where non-linearities from the galileon terms are important, yet quantum effects from terms such as (\ref{quantumaction}) are under control. From the tree-level action containing only the galileon terms (\ref{galileon2}), and where all dimensionful couplings carry the scale $\Lambda$ as appropriate for an effective field theory with cutoff $\Lambda$, we see that the strength of classical nonlinearities is measured by
\be
\alpha_{ cl} \equiv \frac{\di \di \pi}{\Lambda^3} \ ,
\ee
in the sense that the $n$-th order galileon interaction ${\cal L}_n $ is roughly $\alpha_{cl}^{n-2}$ times the kinetic energy for $\pi$.  On the other hand, by factoring out two powers of $\pi$ from the effective action,
\be
\Gamma\sim  \sum_{m'} \big[\alpha_q + \alpha_q^2 +
\alpha_q^3 \log \alpha_q  \big] \di \pi \di \pi \bigg( \frac{\di \di \pi}{\Lambda^3} \bigg)^{m'}  \ ,
\ee
it is clear that the quantity suppressing quantum effects relative to classical ones is 
\be
\alpha_q \equiv \frac{\di^2}{\Lambda^2} \ .
\ee
This separation of scales allows for the existence of regimes in which there exist classical field configurations with non-linearities of order one, $\alpha_{cl}=\di \di \pi / \Lambda^3 \sim 1$, and yet which nevertheless satisfy $\alpha_q\ll 1$, so that quantum effects are under control.   Thus it can be possible to study non-linear classical solutions involving all the gailieon terms, and still trust these solutions in light of quantum corrections\footnote{
In fact, for even larger non-linearities, $\di \di \pi / \Lambda^3 \gg 1$,  quantum fluctuations receive a correspondingly larger kinetic term from the expansion of the non-linear terms about the non-trivial background, thus effectively becoming weakly coupled and suppressing loop corrections even further \cite{Nicolis:2004qq}.}.

An example of such a configuration can be seen in the theory with only the cubic galileon term (setting the others to zero is a technically natural choice, since they are not renormalized) coupled to the trace of the stress tensor of matter, $T$,
\be 
{\cal L} = -3(\partial\pi)^2-{1\over \Lambda^3}(\partial\pi)^2\square\pi+{1\over M_{Pl}}\pi T \ .
\ee
Here $M_{Pl}$ is a mass scale controlling the strength of the coupling to matter (in applications to modified gravity, it is the Planck mass).  

Consider the static spherically symmetric solution, $\pi(r)$, around a point source of mass $M$, $T\sim M\delta^3(r)$ \cite{Nicolis:2004qq}.  The solution transitions, at the distance scale $R_V\equiv{1\over \Lambda}\left(M\over M_{Pl}\right)^{1/3}$, between a linear and non-linear regime,
\be 
\pi(r)\sim \begin{cases} \Lambda^3 R_V^{2} \left(\frac{r}{R_V}\right)^{1/2} & r\ll R_V, \\ \Lambda^3 R_V^2 \left(\frac{R_V}{r}\right) & r\gg R_V. \end{cases} \ 
\ee
Assuming $M\gg M_{Pl}$ so that $R_V\gg {1\over \Lambda}$, we can identify three distinct regimes: Far from the source, at distances $r\gg R_V$, we have $\alpha_{cl}\sim \left(R_V\over r\right)^{3}\ll 1$ and $\alpha_q\sim{1\over \left(r\Lambda\right)^2}\ll 1$, so quantum corrections are under control, but also the interesting classical non-linearities of the cubic term are unimportant.  Close to the source, $r\ll {1\over \Lambda}$, we have $\alpha_{cl}\sim  \left(R_V\over r\right)^{3/2}\gg 1$ and $\alpha_q\sim{1\over \left(r\Lambda\right)^2}\gg 1$.  Here, interesting non-linear effects are important, but quantum effects are not under control, and any attempt to extract physics would require a UV completion.  There is, however, an intermediate range, ${1\over \Lambda}\ll r \ll R_V$, in which $\alpha_{cl}\sim  \left(R_V\over r\right)^{3/2}\gg 1$ and $\alpha_q\sim{1\over \left(r\Lambda\right)^2}\ll 1$ so that interesting non-linear effects are important, while quantum effects are under control.  

An analogous situation is familiar from general relativity.  In that case, the relevant field is the canonically normalized metric perturbation, $g_{\mu\nu}\sim \eta_{\mu\nu}+{1\over M_{Pl}}h_{\mu\nu}$.  The action consists of a linear kinetic term $\sim \partial^2 h^2$, and an infinite number of non-linear terms of the form $\partial^2 h^n$, with $n\geq 3$, which sum up into the Einstein-Hilbert action $\sim M_{Pl}^2 \sqrt{-g}R$.  Diffeomorphism invariance ensures that the relative coefficients of these non-linear terms are not renormalized, so their classical forms can be trusted.  The measure of non-linearity in this case is $\alpha_{cl}\sim h/M_{Pl}$, with non-linear operators suppressed relative to the kinetic terms by powers of this factor.  Quantum effects are expected to generate higher curvature terms, for example $\sqrt{-g}R^2$, ${1\over M_{Pl}^2}\sqrt{-g}R^3$, which will generate higher-derivative operators of the form $\partial^m h^n$, with $m\geq 4$.  These are suppressed relative to classical operators by powers of the factor $\alpha_q\sim{\partial\over M_{Pl}}$.  The analogous spherically symmetric static solution is $h_{\mu\nu}\sim {M\over M_{Pl} r}$, where $M\gg M_{Pl}$ is the total mass of the solution, so that $\alpha_{cl}\sim {M\over M_{Pl}^2 r}$. Therefore, for $r\gg R_S\equiv {M\over M_{Pl}^2}$ (such as in the solar system), classical non-linearities are unimportant, whereas for $r\ll R_S$ (such as inside and near the horizon of a black hole) they dominate.  Since $\alpha_{q}\sim {1\over M_{Pl} r}$, quantum effects are negligible for $r\gg {1\over M_{Pl}}$ but become important near and below the Planck length.  Thus the black hole horizon is the interesting middle regime, where classical non-linearities are large and produce important effects which can be trusted in light of quantum corrections.  These non-linear, quantum-controlled regimes are where interesting models of inflation, cosmology, modified gravity, etc. employing these galileon actions should be placed.

\section{Brane origins of galilean invariance}
\label{braneorigins}

The internal galilean symmetry $\pi\rightarrow \pi+c+b_\mu x^\mu$ of the theories we have discussed above can be thought of as inherited from symmetries of a probe brane floating in a higher dimensional flat bulk, in a small field limit \cite{deRham:2010eu}.  To see this, consider a $3$-brane (3+1 spacetime dimensions) embedded in five dimensional Minkowski space.  Let the bulk coordinates be $X^A$, ranging over $5$ dimensions, and let the brane coordinates be $x^\mu$, ranging over $4$ dimensions.  The bulk metric is flat, $\eta_{AB}$, and the embedding of the brane into the bulk is given by embedding functions $X^A(x)$, which are the dynamical degrees of freedom.

We require the action to be invariant under Poincare transformations of the bulk,
\be 
\label{poincaretransformations} 
\delta_PX^A=\omega^A_{\ B}X^B+\epsilon^A \ ,
\ee
where $\epsilon^A$ and antisymmetric $\omega^A_{\ B}$ are the infinitesimal parameters of the bulk translations and Lorentz transformations respectively.  We also
require the action to be gauge invariant under reparametrizations of the brane,
\be 
\label{gaugetransformations} 
\delta_g X^A=\xi^\mu\partial_\mu X^A \ ,
\ee
where $\xi^\mu(x)$ is the gauge parameter.  

We may use this gauge freedom to fix a unitary gauge
\be 
X^\mu(x)=x^\mu,\ \ \ X^5(x)\equiv\pi (x) \ ,
\ee
where the index set $A$ has been separated into $\mu$ along the brane and $X^5$ transverse to the brane.  Now, 
\be 
\delta_P X^\mu=\omega^\mu_{\ \nu}x^\nu+\omega^\mu_{\ 5}\pi+\epsilon^\mu \ ,
\ee
and so the Poincare transformations~(\ref{poincaretransformations}) do not preserve this gauge. However, the gauge may 
be restored by making a gauge transformation, $\delta_gX^\mu=\xi^\nu\partial_\nu x^\mu=\xi^\mu$ with the choice 
\be 
\xi^\mu=-\omega^\mu_{\ \nu}x^\nu-\omega^\mu_{\ 5}\pi-\epsilon^\mu \ .
\ee
Thus the combined transformation $\delta_{P'}=\delta_P+\delta_g$ leaves the gauge fixing intact and is a symmetry of the gauge fixed action.  Its action on the remaining field $\pi$ is
\be 
\delta_{P'}\pi=-\omega^\mu_{\ \nu}x^\nu\partial_\mu\pi-\epsilon^\mu\partial_\mu \pi+\omega^5_{\ \mu}x^\mu-\omega^\mu_{\ 5}\pi\partial_\mu\pi+\epsilon^5 \ .
\label{combinedtransform}
\ee
The first two terms correspond to unbroken $4$-dimensional Poincare invariance, the second two terms correspond to the broken 
boosts (which will become the galilean symmetry for small $\pi$), and the fifth term is the shift symmetry corresponding to the 
broken translations in the 5th direction.   

In total, the group $ISO(1,4)$ is broken to $ISO(1,3)$.  Renaming $\omega^5_{\ \mu}\equiv \omega_\mu$, and $\epsilon^5\equiv\epsilon$, we 
obtain the internal relativistic invariance under which $\pi$ transforms like a goldstone boson, 
\be 
\delta_{P'}\pi=\omega_{\mu}x^\mu-\omega^\mu\pi\partial_\mu\pi+\epsilon \ .
\ee

This is the relativistic version of the internal galilean invariance we have been considering.  It is the symmetry of theories describing the motion of a brane in a flat bulk, such as DBI.  The non-relativistic limit corresponds to taking the small $\pi$ limit, and in this limit the relativistic invariance reduces to the non-relativistic galilean invariance  
\be 
\label{galileansymmetry} \delta_{P'}\pi=\omega_{\mu}x^\mu+\epsilon \ .
\ee

This co-dimension one construction immediately suggests a generalization.  Consider co-dimension greater than one, so that there will be more than one 
$\pi$ field.  Let the bulk coordinates be $X^A$, ranging over $D$ dimensions, and let the brane coordinates be $x^\mu$, ranging over $d$ dimensions,
so that the co-dimension is $N=D-d$. 
The relevant action will still be invariant under the Poincare transformations (\ref{poincaretransformations}) and the gauge reparameterization symmetries (\ref{gaugetransformations}), and we may use this gauge freedom to fix a unitary gauge
\be 
X^\mu(x)=x^\mu,\ \ \ X^I(x)\equiv\pi^I(x) \ ,
\ee
where the $I$ part of the index $A$ represents directions transverse to the brane.  
Once again the Poincare transformations~(\ref{poincaretransformations}) do not preserve this gauge, since
\be 
\delta_P X^\mu=\omega^\mu_{\ \nu}x^\nu+\omega^\mu_{\ I}\pi^I+\epsilon^\mu \ ,
\ee
but the gauge can be restored by making a gauge transformation, $\delta_gX^\mu=\xi^\nu\partial_\nu x^\mu=\xi^\mu$, with the choice 
\be 
\xi^\mu=-\omega^\mu_{\ \nu}x^\nu-\omega^\mu_{\ I}\pi^I-\epsilon^\mu \ .
\ee
Thus the combined transformation $\delta_{P'}=\delta_P+\delta_g$ leaves the gauge fixing intact and is a symmetry of the gauge fixed action.  Its action on the remaining fields $\pi^I$ is
\be \label{multiinternalpoincare}
\delta_{P'}\pi^I=-\omega^\mu_{\ \nu}x^\nu\partial_\mu\pi^I-\epsilon^\mu\partial_\mu \pi^I+\omega^I_{\ \mu}x^\mu-\omega^\mu_{\ J}\pi^J\partial_\mu\pi^I+\epsilon^I+\omega^I_{\ J}\pi^J \ .
\ee
The first five terms are obvious generalizations of those in~(\ref{combinedtransform}), while the last term is new to co-dimension greater 
than one, and corresponds to the unbroken $SO(N)$ symmetry in the transverse directions.  In total, the group $ISO(1,D-1)$ is broken to $ISO(1,d-1)\times SO(N)$.  

Taking the small $\pi^I$ limit, we find the extended non-relativistic internal galilean invariance under which the $\pi^I$ transform:  
\be
\label{multiinternalgalilean} \delta_{P'}\pi^I=\omega^I_{\ \mu}x^\mu+\epsilon^I+\omega^I_{\ J}\pi^J \ .
\ee
This consists of a galilean invariance acting on each of the $\pi^I$ as in (\ref{generaltermspre}), and, importantly as we shall see, an extra internal $SO(N)$ rotation symmetry under which the $\pi$'s transform as a vector.  

To obtain the multi-field actions invariant under (\ref{multiinternalgalilean}), we must choose the tensor $S$ in (\ref{generaltermspre}) so that it is invariant under $SO(N)$ rotations acting on all its indices.  Equivalently, we must contract up the $I,J,\ldots$ indices on the fields with each other using $\delta_{IJ}$, the only $SO(N)$ invariant tensor (contracting with the epsilon tensor would give a vanishing action).
This simple fact immediately rules out all the lagrangians with an odd number of $\pi$ fields, including the DGP cubic term.  For an even number of $\pi$ fields there are naively two different contractions we can make.  One the one hand, we may contract together the two $\pi$'s appearing with single derivatives, and then the remaining $\pi$'s in any way (the symmetry of $\eta^{\mu_1\nu_1\mu_2\nu_2\cdots\mu_n\nu_n}$ under interchange of $\mu\nu$ pairs with each other makes these all equivalent).  On the other hand, we may contract each of the single derivative $\pi$'s with a double derivative $\pi$. 
By integrating by parts one of the double derivatives in one of the contractions $\partial \pi^I\partial\partial\pi_I$, it is straightforward to show that this second method of contracting the indices is actually equivalent to the first, up to a total derivative. Thus the unique multi-field galileon can be written 
\be 
{\cal L}_{n+1}=n\eta^{\mu_1\nu_1\mu_2\nu_2\cdots\mu_n\nu_n}\left( \partial_{\mu_1}\pi^{I_1}\partial_{\nu_1}\pi_{I_1}\partial_{\mu_2}\partial_{\nu_2}\pi^{I_2} \partial_{\mu_3}\partial_{\nu_3}\pi_{I_2}\cdots\partial_{\mu_{n-1}}\partial_{\nu_{n-1}}\pi^{I_{n-1}}\partial_{\mu_n}\partial_{\nu_n}\pi_{I_{n-1}}\right) \ .
\ee

In four dimensions, there are now therefore only two possible terms; the kinetic term and a fourth order interaction term\footnote{As we were completing the draft of this paper, we received \cite{Padilla:2010ir}, where these exact terms are also considered.},
\bea  
{\cal L}_2&=& \partial_\mu\pi^I\partial^\mu\pi_I, \\
 {\cal L}_4&=&  \partial_\mu\pi^I\partial_\nu \pi_I\left(\partial^\mu\partial_\rho\pi^J\partial^\nu\partial^\rho\pi_J-\partial^\mu\partial^\nu\pi^J\square\pi_J\right)+{1\over 2}  \partial_\mu\pi^I\partial^\mu \pi_I\left(\square\pi^J\square\pi_J-\partial_\nu\partial_\rho\pi^J\partial^\nu\partial^\rho\pi_J\right) \ . 
 \nn 
 \label{multi4thorder}
 \eea
In particular, it is important to note that both the cubic and quintic terms are absent.  
 
This represents an intriguing four dimensional scalar field theory: there is a single possible interaction term, and thus a single free coupling constant  (as in, for example, Yang-Mills theory).  Of course there are other possible terms compatible with the symmetries, namely those which contain two derivatives on every field, and where the field indices are contracted.  However, the quartic term above is the only one with six derivatives and four fields.  All other galilean-invariant terms have at least two derivatives per field.   Thus, as argued in the introduction, there can exist regimes in which the above quartic term is the only one which is important.  Furthermore, as will be shown in section~\ref{nonrenormalizesection}, this term is not renormalized to any order in perturbation theory, so classical calculations in these interesting regimes are in fact exact.  

To fully specify the theory, it is necessary to couple the $\pi$ fields to matter. The simple linear coupling $\pi^I T$, where $T\equiv \eta_{\mu\nu}T^{\mu\nu}$ is the trace of the energy momentum tensor, used in~\cite{Padilla:2010de}, does not respect the $SO(N)$ symmetry of the multi-galileon Lagrangian. There are, of course, many other
couplings that do respect this symmetry. The simplest of these is $\pi^I\pi_I T$, but this has its own drawback, namely that it does not respect the galilean symmetry. To leading order in an expansion in $\pi^I$, a coupling that respects both the internal $SO(N)$ symmetry and the galilean symmetry is given by
\be
\partial_{\mu}\pi^I\partial_{\nu}\pi_I T^{\mu\nu}_{\rm flat} \ ,
\ee
where $ T^{\mu\nu}_{\rm flat}$ is the energy-momentum tensor computed using the flat $4$-dimensional metric $\eta_{\mu\nu}$. Indeed such a coupling will
naturally emerge from a minimal coupling ${\cal L}_{\rm matter}(g_{\mu\nu}, \psi)$ to brane matter $\psi$.

These terms will be important in discussing the phenomenology of multi-galileon theories, but we shall not need to discuss them further in this paper, except for a 
brief comment when we treat quantum corrections in section~\ref{nonrenormalizesection}.

\section{Higher co-dimension branes and actions}

In this section, we show how to construct galilean and internally relativistic invariant scalar field actions from the higher dimensional probe-brane prescription.  This was done in \cite{deRham:2010eu} for the co-dimension one case, and here we extend that approach to higher co-dimension.  

In the co-dimension $1$ case, to obtain an action invariant under the galilean symmetry~(\ref{galileansymmetry}), we need only construct an action for the embedding of a brane $X^A(x),$ which is invariant under the reparametrizations~(\ref{gaugetransformations}) and the Poincare transformations (\ref{poincaretransformations}).  The reparametrizations force the action to be a diffeomorphism scalar constructed out of the induced metric $g_{\mu\nu}\equiv {\partial X^A\over\partial x^\mu} {\partial X^B\over\partial x^\nu} G_{AB}(X)$, where $G_{AB}$ is the bulk metric as a function of the embedding variables $X^A$.  Poincare invariance then requires the bulk metric to be the flat Minkowski metric $G_{AB}(X)=\eta_{AB}$.  Fixing the gauge $X^\mu(x)=x^\mu$ then fixes the induced metric
\be
 g_{\mu\nu}=\eta_{\mu\nu}+\partial_\mu \pi\partial_\nu\pi \ .
\ee
Any action which is a diffeomorphism scalar, evaluated on this metric, will yield an action for $\pi$ having the internal Poincare 
invariance~(\ref{galileansymmetry}), in addition to the usual $4$-dimensional spacetime Poincare invariance.  The ingredients available to construct such an action are the metric $g_{\mu\nu}$, the covariant derivative $\nabla_\mu$ compatible with the induced metric, the Riemann curvature tensor $R^{\rho}_{\ \sigma\mu\nu}$ corresponding to this derivative, and the extrinsic curvature $K_{\mu\nu}$ of the embedding.  Thus, the most general action is
\be
\label{generalaction} 
S=\left. \int d^4x\ \sqrt{-g}F\left(g_{\mu\nu},\nabla_\mu,R^{\rho}_{\ \sigma\mu\nu},K_{\mu\nu}\right)\right|_{g_{\mu\nu}=\eta_{\mu\nu}+\partial_\mu \pi\partial_\nu\pi} \ .
\ee
For example, the DBI action arises from
\be  
\int d^4x\ \sqrt{-g}\rightarrow  \int d^4x\ \sqrt{1+(\partial\pi)^2} \ .
\ee

To recover a galilean-invariant action, with the symmetry~(\ref{galileansymmetry}), we have only to take the small $\pi$ limit.  For example, the DBI action 
above yields the kinetic term ${\cal L}_2$ in this limit.  The DGP cubic term comes from the action $\sim \sqrt{-g}g^{\mu\nu}K_{\mu\nu}$.  Note that this in this construction the brane is merely a probe brane and no de-coupling limit is taken, which is fundamentally different from what occurs in the de-coupling limit of DGP (for the effect of higher order curvature terms in DGP, see for example \cite{Cadoni:2008fb}).  

To generalize this prescription to higher co-dimension, we must now consider diffeomorphism scalars constructed from the induced metric 
\be 
g_{\mu\nu}=\eta_{\mu\nu}+\partial_\mu \pi^I\partial_\nu\pi_I \ .
\ee
A much more difficult question concerns the ingredients from which to construct the action; i.e. the geometric quantities associated with a higher co-dimension brane.  We review the details of how to identify these in Appendix \ref{appendixA}.  The main difference from the co-dimension one
case is that the extrinsic curvature now carries an extra index, $K^i_{\mu\nu}$.  The $i$ index runs over the number of co-dimensions, and is associated with an orthonormal basis in the normal bundle to the hypersurface.  In addition, the covariant derivative $\nabla_\mu$ 
has a connection, $\beta^i_{\mu j}$ that acts on the $i$ index.  For example, the covariant derivative of the extrinsic curvature reads 
\be
\nabla_\rho K^i_{\mu\nu}=\partial_\rho K^i_{\mu\nu}-\Gamma^\sigma_{\rho\mu}K^i_{\sigma\nu}-\Gamma^\sigma_{\rho\nu}K^i_{\mu \sigma}+\beta^i_{\rho j}K^j_{\mu\nu} \ .
\ee
The connection $\beta^i_{\mu j}$ is anti-symmetric in its $i,j$ indices, and so is a new feature appearing in co-dimensions $\geq2$; it vanishes in co-dimension one.  It has an associated curvature, $R^i_{\ j\mu\nu}$.  Therefore, an action of the form
\be
\label{generalmultiaction} 
S=\left. \int d^4x\ \sqrt{-g}F\left(g_{\mu\nu},\nabla_\mu,R^{i}_{\ j\mu\nu},R^{\rho}_{\ \sigma\mu\nu},K^i_{\mu\nu}\right)\right|_{g_{\mu\nu}=\eta_{\mu\nu}+\partial_\mu \pi^I\partial_\nu\pi_I} \ ,
\ee
will have the required relativistic symmetry~(\ref{multiinternalpoincare}), and its small field limit will have the galilean invariance (\ref{multiinternalgalilean}).

\subsection{Brane quantities}

To evaluate the action~(\ref{generalmultiaction}), it is necessary to know how to express the various geometric quantities in terms of the $\pi^I$.  

The tangent vectors to the brane are 
\be 
e^A_{\ \mu}={\partial X^A\over \partial x^\mu}=\begin{cases} \delta^\nu_\mu & A=\nu \ , \\ \partial_\mu\pi^I &A=I \ ,\end{cases}
\ee
and the induced metric is
\be 
g_{\mu\nu}=e^A_{\ \mu}e^B_{\ \nu}\eta_{AB}=\eta_{\mu\nu}+\partial_\mu\pi^I\partial_\nu\pi_I \ ,
\ee
where the $I$ index is raised and lowered with $\delta_{IJ}$.  The inverse metric can then be written as a power series,
\be 
g^{\mu\nu}=\eta^{\mu\nu}-\partial^\mu\pi^I\partial^\nu\pi_I+\mathcal{O}(\pi^4) \ .
\ee

To find the (orthonormal) normal vectors $n^A_{\ i}$ (the index $i$ takes the same values as $I$, but it is the orthonormal frame index, whereas $I$ is the transverse coordinate index), we solve the defining equations
\be 
\label{normalvectorequations} 
e^A_{\ \mu}n^B_{\ i}\eta_{AB}=0,\ \ \ n^A_{\ i}n^B_{\ j}\eta_{AB}=\delta_{ij} \ .
\ee
The first equation tells us that
\be 
n_{Ai}=\begin{cases}-n_{Ii}\partial_\mu\pi^I & A=\mu, \\ n_{Ii} & A=I,\end{cases} \ 
\ee
where $n_{Ii}$ are the as yet undetermined $A=I$ components of $n_{Ai}$.  The second equation of (\ref{normalvectorequations}) then gives
\be 
\label{transversevielbein} 
\delta_{ij}=n^I_{\ i} n^J_{\ j}\left(\partial_\mu \pi_I\partial^\mu\pi_J+\delta_{IJ}\right) \ .
\ee
Thus, the $n^I_{\ i}$ must be chosen to be vielbeins of the transverse ``metric'' $g_{IJ}\equiv\partial_\mu \pi_I\partial^\mu\pi_J+\delta_{IJ}$.  The ambiguity in this choice due to local $O(N)$ transformations reflects the freedom to change orthonormal basis in the normal space of the brane.   The vielbeins summed over their Lorentz indices $i,j$ give the inverse of the metric to $g_{IJ}$, which expanded in powers of $\pi$ gives
\be 
n^I_{\ i}n^{J}_{\ j}\delta^{ij}=\delta^{IJ}-\partial_\mu\pi^I\partial^\mu\pi^J+\mathcal{O}(\pi^4) \ .
\ee
The metric determinant can be expanded as
\be 
\sqrt{-g}=1+\half \partial_\mu\pi^I\partial^\mu\pi_I+\mathcal{O}(\pi^4) \ ,
\ee
and the extrinsic curvature is 
\bea 
K_{i\mu\nu}&=&e^A_{\ \mu}e^B_{\ \nu}\nabla_An_{Bi}=e^B_{\ \nu}\partial_\mu n_{Bi}  \nn \\
&=&\partial_\mu n_{\nu i}+\partial_\nu\pi^I\partial_\mu n_{Ii}=-\partial_\mu \left(n_{Ii}\partial_\nu\pi^I\right)+\partial_\nu\pi^I\partial_\mu n_{Ii} \nn \\
&=&-n_{Ii}\partial_\mu\partial_\nu \pi^I \ .
\eea

Finally, the twist connection is 
\bea
\beta_{\mu ij}=n^B_{\ i}e^A_{\ \mu}\nabla_A n_{Bj}&=&n^B_{\ i}\partial_\mu n_{Bj}=n^\nu_{\ i}\partial_\mu n_{\nu j}+n^I_{\ i}\partial_\mu n_{Ij} \nn \\
&=&\partial^\nu \pi^In_{Ii}\partial_\mu\left(\partial_\nu \pi^J n_{Jj}\right)+n^I_{\ i}\partial_\mu n_{Ij} \nn \\
&=& \left(\delta^{IJ}+\partial_\nu \pi^I \partial^\nu \pi^J\right)n_{Ii}\partial_\mu n_{Jj}+n_{Ii}n_{Jj}\partial^\nu \pi^I\partial_\mu\partial_\nu\pi^J \ .
\eea

The action~(\ref{generalmultiaction}) is an $SO(N)$ scalar, and so will not depend on how the $\eta^I_{\ i}$ are chosen.

\subsection{Lovelock terms and the probe brane prescription}

A general choice for the action~(\ref{generalmultiaction}) will not lead to scalar field equations that are second order.  One of the key insights of de Rham and Tolley \cite{deRham:2010eu} is that the actions that do lead to second order equations are precisely those that are related to Lovelock invariants.   It is well-known that the possible
extensions of Einstein gravity which remain second order are given by the famous Lovelock terms
\cite{Lovelock:1971yv}.  These terms are combinations of powers of the Riemann tensor which are dimensional continuations of characteristic classes.  We summarize some properties of these terms in Appendix \ref{lovelockappendix}.  The problem of finding extensions of the
$\pi$ Lagrangian which possess second-order equations of motion is therefore
equivalent to the problem of finding extensions of higher-dimensional 
Einstein gravity which have second-order equations of motion.  

In the presence of lower-dimensional hypersurfaces or branes, 
Lovelock gravity in the bulk must be supplemented by terms which depend on the intrinsic and extrinsic geometry of the
brane.   These additional surface terms are required in order to ensure that the
variational problem of the combined brane/bulk system is well posed \cite{Dyer:2008hb}.  The variation of the surface term precisely cancels the 
higher-derivative variations on the surface which would otherwise appear in the equations of
motion.
For the case of Einstein gravity these considerations lead one to supplement
the Einstein-Hilbert lagrangian by the Gibbons-Hawking-York boundary term
\cite{Gibbons:1976ue,York:1972sj}
\begin{equation}\label{e:EHGH}
S = \int_M  \dd^4 x \ \sqrt{-g}R \ + 
2 \int \ \dd^3 y\sqrt{-h}K \, ,
\end{equation}
where $x,y$ are the bulk and brane coordinates respectively, $R$ is the Ricci scalar of
the bulk metric $g$, and $K$ is the trace of the extrinsic curvature of the
induced metric $h$ on the brane.

The addition of Gibbons-Hawking-York boundary terms is closely related to the
issue of matching conditions for the bulk metric.  When there are
distributional sources of stress-energy supported on the brane, the 
extrinsic curvatures on either side of the brane must be related to the brane
stress energy in a specific way.  This relationship can be derived by 
supplementing (\ref{e:EHGH}) by an action for the brane matter, then varying
with respect to the bulk and induced metrics.

Similarly, boundary terms (Myers terms) for the Lovelock invariants must be added \cite{Myers:1987yn,Miskovic:2007mg}.  The prescription of \cite{deRham:2010eu} is as follows:  the $d$-dimensional single field galileon terms with an even number $N$ of $\pi$'s are obtained from the $(N-2)$-th Lovelock term on the brane, constructed from the brane metric (see Appendix \ref{lovelockappendix} for numbering convention of the Lovelock terms).  The terms with an odd number $N$ of $\pi$'s are obtained from the boundary term of the $(N-1)$-th $d+1$ dimensional bulk Lovelock term.  For instance, in $d=4$, the kinetic term with two $\pi$'s is obtained from $\sqrt{-g}$ on the brane; the cubic $\pi$ term is obtained from the Gibbons-Hawking-York term $\sqrt{-g}K$; the quartic term is obtained from $\sqrt{-g}R$; and the quintic term arises from the boundary term of the bulk Gauss-Bonnet invariant.  There are no further non-trivial Lovelock terms for $d=4$, in either the brane or the bulk, corresponding to the fact that there are no further non-trivial galileon terms.

Our goal is to build upon this prescription, and extend it to higher co-dimension.  For this, we need the corresponding higher-co-dimension boundary terms induced by the bulk Lovelock invariants.   These were studied by Charmousis and Zegers~\cite{Charmousis:2005ey}, who found that, despite
the freedom to specify a fairly general bulk gravitational theory and
number of extra dimensions, the resulting four-dimensional 
terms are surprisingly constrained, corresponding to the fact that the multi-galileon action is essentially unique.

The summary of brane terms claimed in \cite{Charmousis:2005ey}, for a brane of dimension $d=4$, is as follows:

\begin{itemize}

\item If the co-dimension $N$ is odd and $N\ne 3$,  one obtains the dimensional continuation of the Gibbons-Hawking-York and Myers terms, with the extrinsic curvature replaced by a distinguished normal component of $K^i_{\mu\nu}$.  
When $N=3$, there are additional terms involving the extrinsic curvature and
the boundary term is not the dimensional continuation of the Myers term.

\item If $N$ is even (see also \cite{Charmousis:2005ez}),

\begin{itemize}

\item If $N=2$, then the boundary terms include only 
 a brane cosmological constant,
and the following term
\be\label{e:Neven2}
{\cal L}_{N=2} =
\sqrt{-g} \left(R[g] - (K^i)^2 + K_{\mu\nu}^i K^{\mu\nu}_{i}
 \right).
\ee

\item If $N>2$, the boundary term includes only a brane cosmological constant
and an induced Einstein-Hilbert term.  

\end{itemize}

\end{itemize}

In what follows, we will restrict to the even co-dimension case, since it is unclear to us how the normal components in the odd terms are to be interpreted.

\subsection{Recovering the multi-field galileon}

As we saw in the previous subsection, the unique brane action in four dimensions for even co-dimension $\geq 4$ is
\be 
S=\int d^4 x\ \sqrt{-g}\left(-a_2+a_4 R\right) \ .
\ee
The galileon action is obtained by substituting $g_{\mu\nu}=\eta_{\mu\nu}+\partial_\mu\pi^I\partial_\nu\pi_I$, and expanding each term to lowest non-trivial order in $\pi$.  The cosmological constant term yields an $\mathcal{O}(\pi^2)$ piece, and the Einstein-Hilbert term yields an $\mathcal{O}(\pi^4)$ piece.  Up to total derivatives, we have\footnote{A nice way to expand the Einstein-Hilbert term is to think in terms of a metric perturbation, $g_{\mu\nu}=\eta_{\mu\nu}+h_{\mu\nu}$, where $h_{\mu\nu}=\partial_\mu\pi^I\partial_\nu\pi_I$, as in weak-field studies of general relativity.  Then fourth order in $\pi$ is second order in $h_{\mu\nu}$, but the second order in $h_{\mu\nu}$ is just the familiar lagrangian for a massless graviton, \[\half\delta^2\left(\sqrt{-g}R\right)=-\frac{1}{4}\partial_\lambda h_{\mu\nu}\partial^\lambda h^{\mu\nu}+\half\partial_\mu h_{\nu\lambda}\partial^\nu h^{\mu\lambda}-\half\partial_\mu h^{\mu\nu}\partial_\nu h+\frac{1}{4}\partial_\lambda h\partial^\lambda h+(\rm total\ derivative).\]  Evaluating this on $h_{\mu\nu}=\partial_\mu\pi^I\partial_\nu\pi_I $ gives (apart from the total derivative) the coefficient of $a_4$ in~(\ref{fourthorderaction}).}
\be 
\label{fourthorderaction} 
S=\int d^4 x\ \left[-a_2 \ \half\partial_\mu\pi^I\partial^\mu\pi_I  +a_4\ \partial_\mu\pi^I\partial_\nu\pi^J\left(\partial_\lambda\partial^\mu\pi_J\partial^\lambda\partial^\nu\pi_I-\partial^\mu\partial^\nu\pi_I\square\pi_J \right)\right] \ .
\ee
Again, by adding a total derivative, we can see that the $a_4$ term is proportional to the fourth order term~(\ref{multi4thorder}), so we recover the four dimensional multi-field galileon model,
\be 
S=\int d^4 x\ \left[-{1\over 2} a_2  {\cal L}_2+{1\over 2} a_4  {\cal L}_4\right] \ .
\ee

The equations of motion are
\bea 
{\delta S\over \delta \pi^I}= && a_2\square \pi_I \nn \\ 
+&&a_4\left[\square\pi_I \left(\partial_\mu\partial_\nu\pi_J\partial^\mu\partial^\nu\pi^J- \square\pi^J\square\pi_J\right)+2\partial_\mu\partial_\nu\pi_I \left(  \partial^\mu\partial^\nu\pi_J\square\pi^J -\partial^\mu\partial_\lambda\pi_J\partial^\nu\partial^\lambda\pi^J\right)\right] \ . \nn \\
\label{multi4eom}
\eea

For co-dimension two, there is the additional $K^2$ part to the boundary term.  This cancels the contribution from the Ricci scalar, and thus yields nothing new.  Therefore,~(\ref{fourthorderaction}) is the unique multi-galileon term in four dimensions and any even co-dimension.  Keeping all orders in $\pi$ would lead to a relativistically invariant action, a multi-field generalization of DBI with second order equations.

\section{De-Sitter solutions of the unique 4-th order action}

While the main aim of this paper is a derivation of the unique multi-galileon action and its origin in the geometry of braneworlds in
co-dimension greater than one, it is worth exploring the simplest properties of the resulting theories. Perhaps the most straightforward question to ask concerns
the nature of maximally symmetric solutions to the equations of motion.  If the galileon were being used to describe a modification to gravity, the interest would be in scalar field profiles that correspond to a gravitational de Sitter background solution.  As was argued in \cite{Nicolis:2004qq,Nicolis:2008in} for the single field galileons, these profiles take the form $\sim x^\mu x_\mu$ at short distances, where $x^\mu$ is the spacetime coordinate.  In fact, this is easy to see geometrically; a de Sitter 3-brane can be embedded in $5$-dimensional Minkowski space via the equation $X^A X_A={\cal R}^2$, where ${\cal R}$ is the radius of the de Sitter space.  Thus, taking $x^\mu=X^\mu$ as the brane coordinates and $y=X^5$ as the transverse coordinate, the $\pi$ profile is
\be
\pi\sim y=\sqrt {{\cal R}^2-x^\mu x_\mu}\approx {-1\over 2{\cal R}}x^\mu x_\mu+{\rm constant} \ , 
\ee
where we have expanded for short distances.   The constant can be ignored due to the shift symmetry of $\pi$.  

Thus we consider the ansatz  
 \be \pi^I=\Lambda^I x^\mu x_{\mu},\ee
 where $\Lambda^I$ are constants. This corresponds to a de Sitter brane bending along some general transverse direction.  It is easy to see that (\ref{multi4eom}) then
yields the condition 
\be 
a_2\Lambda^I-24 a_4\Lambda^I \Lambda^2=0 \ ,
\ee
where $\Lambda^2\equiv \Lambda^I\Lambda_I$.  A non-trivial solution requires setting
\be 
\Lambda^2={a_2\over 24 a_4} \ ,
\ee 
and exists if and only if $a_2$ and $a_4$ have the same sign.

To study the stability of these solutions, we expand the field in fluctuations about the de Sitter solution, setting $\pi^I=\Lambda^I x^\mu x_{\mu}+\delta\pi^I.$
The part of the action quadratic in fluctuations reads
\be {\cal L}_{{\cal O}(\delta\pi^2)}=48a_4\Lambda_I\Lambda_J\partial_\mu\pi^I\partial^\mu\pi^J.\ee
Since $\Lambda_I\Lambda_J$ is a matrix of rank 1, only one of the $\pi$ fields propagates on this background. No new degrees of freedom appear (contrary to the situation for example in massive gravity, where a sixth degree of freedom appears around non-trivial backgrounds).  This is a general feature of galileon-type theories - the second order property of the equations guarantee that no new degrees of freedom propagate around non-trivial backgrounds.

However, since $\Lambda_I\Lambda_J$ is a positive matrix, our degree of freedom is a ghost if $a_4>0$, signaling that this solution is unstable\footnote{Note that we use the $(-,+,+,+)$ metric convention.}.  If $a_2>0$, so that there is no ghost around flat space, then we must have $a_4>0$ for a non-trivial de Sitter solution to exist, and hence there will be a ghost around the de Sitter solution.  If we choose $a_4<0$ to avoid the ghost around de Sitter, then we necessarily have $a_2<0$ and the ghost reappears around flat space.

\section{\label{nonrenormalizesection}Quantum properties and non-renormalization}

One of the most interesting properties of the galileon actions is their stability under quantum corrections (discussed for the special case of a single field cubic term in \cite{Luty:2003vm}).  In this section, we show that, in any theory with galilean symmetry on each field, the general multi-field scalar galileon term receives no quantum corrections, to any order in perturbation theory, in any number of dimensions.

Consider an effective field theory for scalars $\pi^I$ invariant under individual galilean transformations $\pi^I\rightarrow \pi^I+c^I+b^I_\mu x^\mu$ (in this section we remain more general and do not impose any additional internal symmetries among the $\pi$ fields).  The classical action may contain the general multi-field scalar galileon  terms (\ref{generaltermspre}), 
\be 
\label{generalterms} 
{\cal L}_{n+1}\sim S_{I_1I_2\cdots I_{n+1}}\eta^{\mu_1\nu_1\mu_2\nu_2\cdots\mu_n\nu_n}\left(\pi^{I_{n+1}} \partial_{\mu_1}\partial_{\nu_1}\pi^{I_1}\partial_{\mu_2}\partial_{\nu_2}\pi^{I_2}\cdots\partial_{\mu_n}\partial_{\nu_n}\pi^{I_{n}}\right) \ ,
\ee
with $S_{I_1I_2\cdots I_{n+1}}$ a symmetric constant tensor.  These are the only terms that yield second order equations of motion, and are the only $n$-field terms that contain $2n-2$ derivatives.  There are no terms with $n$ fields that contain fewer that $2n-2$ derivatives, but there are plenty of possible galilean invariant terms with $\geq 2n$ derivatives (i.e. any term with two or more derivatives on each $\pi$), and we also allow for the presence of these terms in the classical action.

Consider quantum corrections by calculating the quantum effective action for the classical field, $\Gamma(\pi^c)$, expanded about the expectation value $\langle\pi\rangle=0$,
\be
\Gamma(\pi^c)= \Gamma^{(2)}\pi^c\pi^c+ \Gamma^{(3)}\pi^c\pi^c\pi^c+\cdots \ .
\ee  
The term $\Gamma^{(n)}$ is calculated in momentum space by summing all $1PI$ diagrams with $n$ external $\pi$ lines.  The position space action is obtained by expanding in powers of the external momenta, and then replacing the momenta with derivatives.  $\Gamma^{(n)}$ thus contains all terms with $n$-fields and any number of derivatives, the number of derivatives being the power of external momenta in the expansion of the $n$-point $1PI$ diagram.

To show that the terms (\ref{generalterms}) do not receive quantum corrections, we argue that all $n$ point diagrams, constructed with vertices drawn from the classical action, contain at least $2n$ powers of the external momenta.  To do this, we show that each external line contributes at least two powers of the external momenta.  

\begin{figure}[h!]
\begin{center}
\epsfig{file=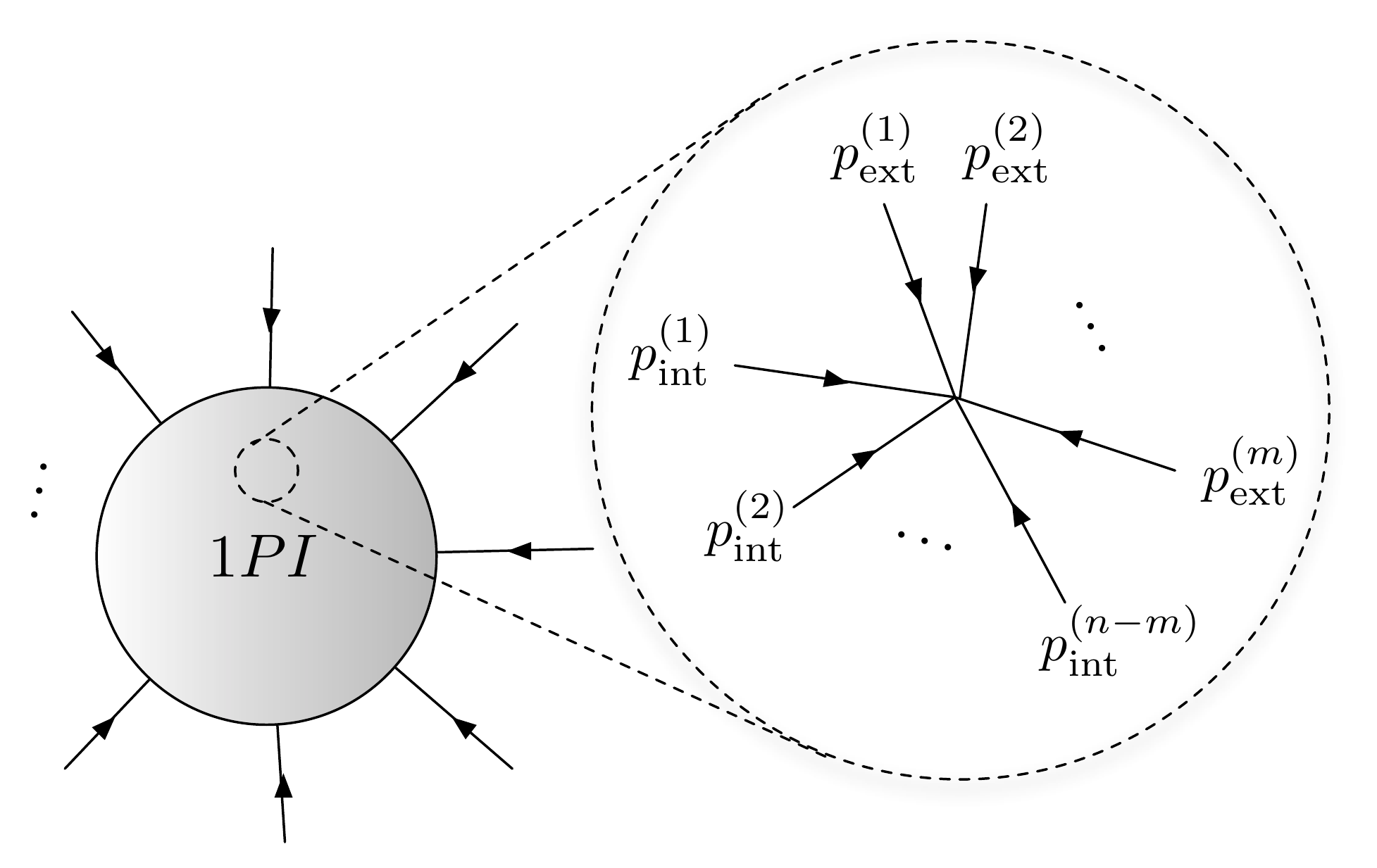, width=10cm}
\caption{A general Feynman diagram and vertex potentially contributing to quantum corrections to the galileon terms. As we prove, such corrections
vanish in these theories.}
\label{feynman}
\end{center}
\end{figure}

Focus on any given vertex connected to external lines, as depicted in figure~\ref{feynman}. 
If the external lines hit only the $\partial\partial\pi$ pieces (this encompases the case where the vertex is drawn from non-galileon terms, i.e. terms with at least two derivatives on every $\pi$), then the vertex will contribute two powers of momentum for each external line.  The other possibility is that one of the external lines hits the undifferentiated $\pi$ in a vertex of the form (\ref{generalterms}).  Suppose there are $m$ external lines, then the contraction looks like 
\be {\cal L}_{n+1}\sim S_{I_1I_2\cdots I_{n+1}}\eta^{\mu_1\nu_1\mu_2\nu_2\cdots\mu_n\nu_n}\left(\pi^{I_{n+1}}_{\rm ext} \partial_{\mu_1}\partial_{\nu_1}\pi^{I_1}_{\rm ext}\cdots\partial_{\mu_{m-1}}\partial_{\nu_{m-1}}\pi^{I_{m-1}}_{\rm ext}\partial_{\mu_{m}}\partial_{\nu_{m}}\pi^{I_{m}}_{\rm int}\cdots\partial_{\mu_n}\partial_{\nu_n}\pi^{I_{n}}_{\rm int}\right).
\ee
Using the antisymmetry of $\eta$, we may write the part containing $\pi_{\rm int}$ as a double total derivative,
\be  
{\cal L}_{n+1}\sim S_{I_1I_2\cdots I_{n+1}}\eta^{\mu_1\nu_1\mu_2\nu_2\cdots\mu_n\nu_n}\left(\pi^{I_{n+1}}_{\rm ext} \partial_{\mu_1}\partial_{\nu_1}\pi^{I_1}_{\rm ext}\cdots\partial_{\mu_{m-1}}\partial_{\nu_{m-1}}\pi^{I_{m-1}}_{\rm ext}\partial_{\mu_{m}}\partial_{\nu_{m}}\left[\pi^{I_{m}}_{\rm int}\cdots\partial_{\mu_n}\partial_{\nu_n}\pi^{I_{n}}_{\rm int}\right]\right).
\ee
The Feynman rule for this contraction therefore contains two factors of the sum of the internal momenta, $\sum p_{\rm int}$.  By momentum conservation at each vertex, we can trade these for the external momenta, $-\sum p_{\rm ext}$.  This adds two powers of $p_{\rm ext}$ to the count, making up for the undifferentiated $\pi$, and bringing the total to $2n$.    

This means that the expansion of the $n$-point diagram in powers of external momenta must start at order $\geq 2n$, so the terms of the form (\ref{generalterms}), which have $2n-2$ derivatives, cannot receive new contributions.  This holds at all loops in perturbation theory, and regardless of any other terms of the form $(\partial\partial\pi)^{\rm power}$ that are present in the classical action.  Note that the kinetic term is of the form (\ref{generalterms}), so there is no wavefunction renormalization in these theories.

This non-renormalization theorem is not a consequence of a symmetry of the theories.  In quantum field theory, we are used to seeing terms vanish or stay naturally small because of symmetry, but here the terms (\ref{generalterms}) are compatible with the symmetries and yet still do not receive quantum corrections.  The situation is more analogous to that in supersymmetric theories, where superpotentials do not receive quantum corrections even though they are compatible with supersymmetry.  In the supersymmetric case there is an underlying reason, namely holomorphy of the superpotential.  Here, the reason seems to be that the galileon terms just do not contain sufficient numbers of derivatives, yet still manage to be galilean-invariant.  

These conclusions may be changed when couplings to matter, as mentioned in section~\ref{braneorigins}, are included. However, any corrections to the galileon
terms must be proportional to the $\pi$-matter coupling, and thus must go to zero as these couplings do. In particular, in applications to modified gravity,
couplings to matter will typically be Planck-suppressed.

\section{Conclusions}
Braneworld models with induced gravity have been extensively studied in co-dimension one. 
The relevant action contains a nonlinear cubic interaction which yields
interesting cosmological phenomenology and strict constraints from local tests of gravity. In this paper we have systematically
extended this idea to higher co-dimension, and have explored the origin of the allowed terms, and the symmetry group under which they transform, in the geometric terms arising
in the action for the brane in the higher dimensional space. The relevant terms are generalizations of those obtained in~\cite{deRham:2010eu} and
are related to the bulk Lovelock terms and their associated boundary actions.

The existence of more than one extra spatial dimension allows for multiple brane bending modes and
correspondingly the $4$-dimensional effective theory contains multiple galileon fields. 
Interestingly, the residual symmetry group of this theory contains an internal $SO(N)$ subgroup that forbids nonlinear interactions 
with odd numbers of galileon fields. Thus, the usual galileon term does not remain in higher co-dimension. Instead what results
is a highly constrained theory with a single coupling constant, governing the strength of a unique nonlinear quartic derivative interaction. We
have further proved a general non-renormalization theorem, which demonstrates that in any number of co-dimensions, the resulting galileon
theory contains only terms that receive no quantum corrections at any loop in perturbation theory.

Multi-galileon theories in principle possess a rich and interesting phenomenology. While not the main thrust of this paper, we have initiated
such a study by considering the simplest example of maximally symmetric backgrounds. For suitable choices of signs of the coupling
constants, we have demonstrated the existence of a de Sitter background, and have explored the stability of the theory around it. The 
result is a generalization of the familiar DGP case of a ghost in the accelerating branch. More precisely, we demonstrate that when the de
Sitter solution exists, then it is possible for either it, or the flat space solution to be ghost-free, but not both. The implications of this
result for self-accelerating cosmologies from multi-galileon theories remain to be seen.

\bigskip
\goodbreak
\centerline{\bf Acknowledgements}
\noindent
\\
The authors are grateful to Melinda Andrews, Christos Charmousis, Lam Hui, Justin Khoury and Alberto Nicolis for discussions. This work is supported in part by NASA ATP grant NNX08AH27G, NSF grant PHY-0930521, and by Department of Energy grant DE-FG05-95ER40893-A020. MT is also supported by the Fay R. and Eugene L. Langberg chair.

\appendix

\section{\label{appendixA}Mathematics of higher co-dimension hypersurfaces}

\begin{figure}[h!]
\begin{center}
\epsfig{file=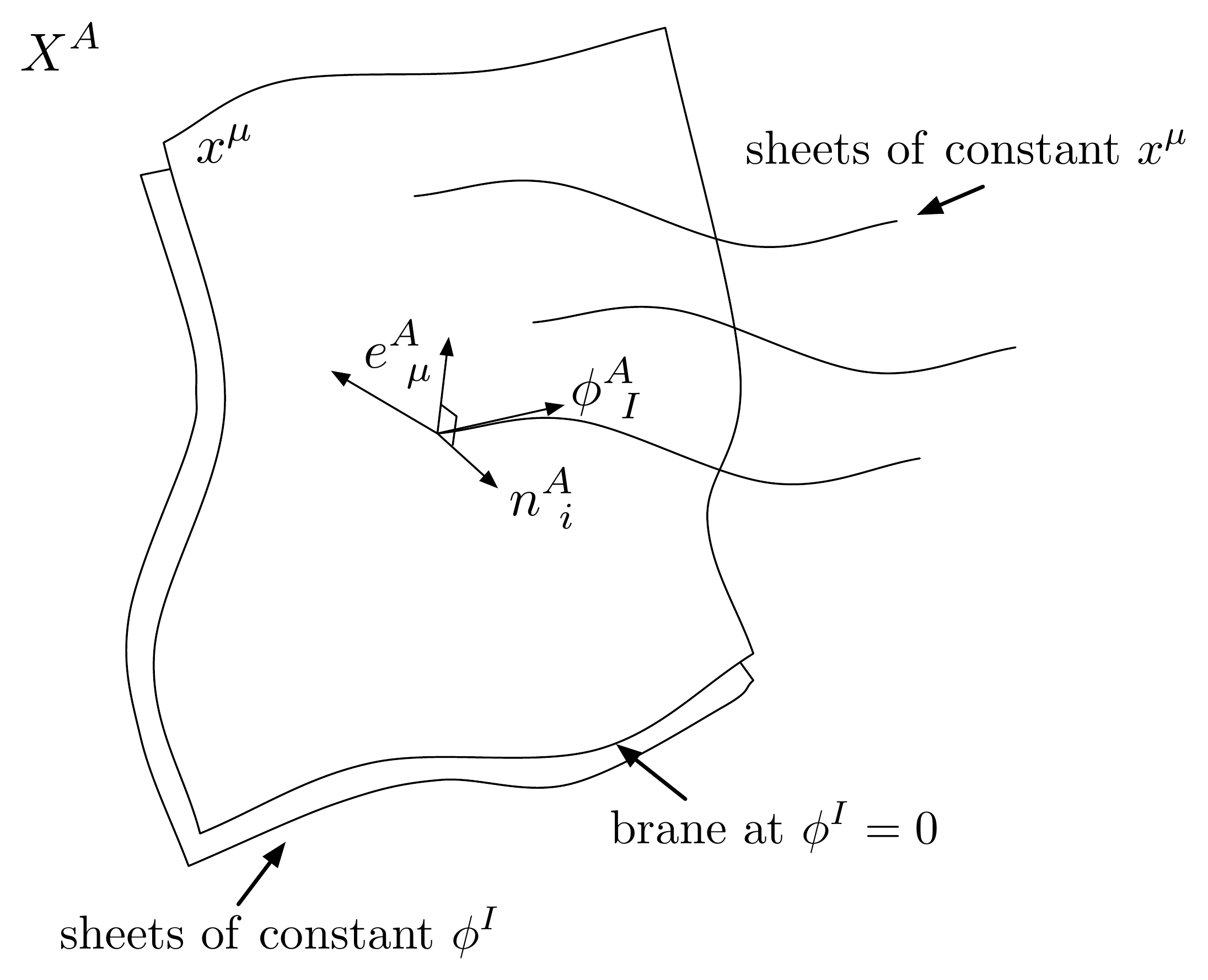,height=3.5in,width=4.0in}
\caption{The geometric setup for a higher co-dimension brane}
\label{brane}
\end{center}
\end{figure}

Here we describe the formalism necessary to deal with submanifolds of higher co-dimension.  The geometric setup is shown in figure~\ref{brane}.

\subsection{Submanifolds and adapted basis}

Let ${\cal M}$ be a manifold of dimension $D$, with coordinates $X^A$.  We describe an $d$-dimensional submanifold ${\cal N}$ of ${\cal M}$ as the locus of zeros of $N\equiv D-d$ functions
\be 
\phi^I(X)=0,\ \ \ I=1\ldots N \ .
\ee
The level sets of $\phi^I$ give a foliation of ${\cal M}$ into a family of $d$-dimensional submanifolds, of which ${\cal N}$ is a member.  The submanifolds have co-dimension $N$.

We now describe a new set of coordinates on ${\cal M}$, adapted to the foliation.  First, set up coordinates $x^\mu$, $\mu=1\ldots d$, on ${\cal N}$.   Now set up functions $x^\mu(X)$ which are independent of the $\phi^I(X)$ and each other, and whose values on ${\cal N}$ coincide with the coordinates $x$ on ${\cal N}$. The level sets of the $x^\mu(X)$ will define a congruence of curves intersecting all the submanifolds.  We use this congruence to assign coordinates on all the other submanifolds from those on ${\cal N}$, so that the coordinates are given by $x^\mu$.  The $x^\mu$ along with the $\phi^I$ now form a new coordinate system on ${\cal M}$.  We have a transformation from these new coordinates to the old coordinates $X^A$,  
\be 
X^A(x^\mu,\phi^I),\ \ \phi^I(X^A),\ x^\mu(X^A) \ .
\ee

The basis vectors of this new coordinate system are
\be 
\phi^A_{\ I}={\partial X^A\over \partial\phi^I},\ \ \ e^A_{\ \mu}={\partial X^A\over \partial x^\mu} \ .
\ee
The basis one forms are 
\be 
\phi^{\ I}_A={\partial \phi^I\over \partial X^A},\ \ \ \tilde{e}^{\ \mu}_A={\partial x^\mu \over \partial X^A} \ .
\ee
(We have put a tilde on $\tilde{e}_A^{\ \mu}$, because later we will introduce a metric and use normal vectors in place of $\phi^A_{\ I}$, so the dual basis will have to change, at which point we'll use $ {e}_A^{\ \mu}$.)

They satisfy duality and completeness relations
\be 
\phi^A_{\ I}\phi^{\ J}_A=\delta_I^J,\ \ e^A_{\ \mu}\tilde{e}^{\ \nu}_A=\delta_\mu^\nu,\ \ \phi^A_{\ I}\tilde{e}^{\ \mu}_A=e^A_{\ \mu}\phi^{\ I}_A=0 \ .
\ee
\be 
\phi^A_{\ I}\phi^{\ I}_B+e^A_{\ \mu}\tilde{e}^{\ \mu}_B=\delta^A_{\ B} \ .
\ee

\subsection{Metric}

Now suppose there is a bulk metric $G_{AB}$.  The metric can have any signature, but we demand that the foliation be non-null.  There is now a well defined normal subspace of the tangent space of ${\cal M}$ at each point, which may be different from the subspace defined by the congruence, which is spanned by $\phi^A_{\ I}$.  We set up a basis consisting of $N$ orthonormal normal vectors $n^A_{\ i}$, as well as the $e^A_{\ \mu}$ which are not required to be orthonormal among themselves.  
\be 
G_{AB}n^A_{\ i}n^B_{\ j}=\eta_{ij},\ \ G_{AB}e^A_{\ a}n^B_{\ j}=0 \ . 
\ee
Here $\eta_{ij}$ is the $N$-dimensional flat Minkowski or euclidean metric carrying whatever signature the transverse space has.   
 We define the associated dual forms ${e}^{\ \mu}_A$, $n_A^{\ i}$, at each point, 
\be 
n^A_{\ i}n^{\ j}_A=\delta_i^j,\ \ e^A_{\ \nu}{e}^{\ \mu}_A=\delta^\mu_\nu,\ \ n^A_{\ i}{e}^{\ \mu}_A=e^A_{\ \mu}n^{\ i}_A=0 \ .
\ee
\be 
n^A_{\ i}n^{\ i}_B+e^A_{\ \mu}{e}^{\ \mu}_B=\delta^A_{\ B} \ .
\ee
This choice of basis is unique up to local orthogonal rotations in the normal space.  

\subsection{Parallel and normal tensors}

First we consider tensors which are parallel to the submanifold ${\cal N}$.  A vector $V^A$ is parallel if it admits the decomposition $V^A=V^\mu e^A_{\ \mu}$.  A form $V_A$ is parallel if it admits the decomposition $V_A=V_\mu {e}_A^{\ \mu}$.  (Notice that, unlike a vector, the notion of a form being parallel depends on the dual basis, will change if the dual basis is changed, and hence depends on the metric.)  Similarly, a general tensor $T^{AB\ldots}_{\ \ C\ldots}$ is parallel if it admits an analogous decomposition, 
\be 
T^{AB\ldots}_{\ \ C\ldots}=A^{\mu\nu\ldots}_{\ \ \rho\ldots}e^A_{\ \mu}e^B_{\ \nu} {e}_C^{\ \rho}\cdots \ .
\ee
There is a bijective relation between tensors on the submanifold ${\cal N}$ (really a $N$-parameter family of tensors, one on each surface, parametrized by $\phi^I$) and parallel tensors in the bulk.  Given a parallel bulk tensor $T^{AB\ldots}_{\ \ C\ldots}$, it corresponds to the submanifold tensor $A^{\mu\nu\ldots}_{\ \ \rho\ldots}$, and vice versa.  

Define the projection tensor
\be 
{P}^A_{\ B}\equiv\delta^A_{B}-n^A_{\ i}n^{\ i}_{B} \ .
\ee
It projects the tangent space of ${\cal M}$ onto the tangent space of ${\cal N}$, along the subspace spanned by $n^A_{\ i}$.  It satisfies 
\bea 
{P}^A_{\ C} {P}^C_{\ B}={P}^A_{\ B},\\ {P}^A_{\ B}e^B_{\ \mu}=e^A_{\ \mu},\ \ {P}^A_{\ B}n^B_{\ i}=0, \\ {P}^A_{\ B}{e}_A^{\ \mu}={e}_B^{\ \mu}, \ \ P^A_{\ B}n_A^{\ i}=0 \ .
\eea

Given any bulk tensor, $T^{AB\ldots}_{\ \ C\ldots}$, we can make a parallel tensor by projecting it along all its indices,
\be 
T^{\parallel AB\ldots}_{\ \ \ C\ldots}\equiv {P}^A_{\ D}{P}^B_{\ E}{P}^F_{\ C} \cdots T^{DE\ldots}_{\ \ F\ldots} \ .
\ee
A tensor is parallel if and only if it is equal to its projection.  

We have the relation
\be 
e^A_{\ \mu}{e}^{\ \mu}_{B}={P}^A_{\ B} \ .
\ee  

Projecting the metric gives the induced metric $h_{AB}$ on the hypersurfaces, whose intrinsic components we denote $g_{\mu\nu}$, 
\be 
h_{AB}=P^C_{\ A}P^D_{\ B}G_{CD}=g_{\mu\nu}e_A^{\ \mu}e_B^{\ \nu},\ \ \ g_{\mu\nu}=e^A_{\ \mu}e^B_{\ \nu}h_{AB}=e^A_{\ \mu}e^B_{\ \nu}g_{AB} \ .
\ee
We raise and lower bulk indices $A,B,\ldots$ with $G_{AB}$ and its inverse $G^{AB}$, and we raise and lower submanifold indices $\mu,\nu,\ldots$ with $g_{\mu\nu}$ and its inverse $g^{\mu\nu}$.  We raise and lower perpendicular indices $i,j,\dots$ using $\eta_{ij}$ and its inverse $\eta^{ij}$.  In particular, we have, 
\bea 
g^{\mu\nu}G_{AB}e^B_{\ \nu}&=&e_A^{\ \mu},\ \ \ \eta^{ij}G_{AB}n^B_{\ j}=n_A^{\ i},\\
g_{\mu\nu}G^{AB}e_B^{\ \nu}&=&e^A_{\ \mu},\ \ \ \eta_{ij}G^{AB}n_B^{\ j}=n^A_{\ i} \ .
\eea
as well as
\be 
G_{AC}P^C_{\ B}=h_{AB},\ \ \  G^{AC}P^B_{\ C}=h^{AB} \ .
\ee

We next consider tensors which are normal to the submanifolds.  A vector $V^A$ is normal if it admits the decomposition $V^A=V^i n^A_{\ i}$.  A form $V_A$ is normal if it admits the decomposition $V_A=V_i n_A^{\ i}$.  Similarly, a general tensor $T^{AB\ldots}_{\ \ C\ldots}$ is normal if it admits an analogous decomposition, 
\be 
T^{AB\ldots}_{\ \ C\ldots}=A^{ij\ldots}_{\ \ k\ldots}n^A_{\ i}n^B_{\ j}n_C^{\ k}\cdots \ .
\ee

Define another projection tensor
\be 
P^A_{\perp B}\equiv \delta^A_{B}-e^A_{\ \mu}e^{\ \mu}_{B} \ .
\ee
It projects the tangent space of ${\cal M}$ onto the normal space of ${\cal N}$, along the tangent space.  It satisfies 
\bea 
P^A_{\perp C}P^C_{\perp B}=P^A_{\perp B},\\  P^A_{\perp B}n^B_{\ i}=n^A_{\ i},\ \ P^A_{\perp B}e^B_{\ \mu}=0 \\ P^A_{\perp B}n_A^{\ i}=n_B^{\ i},\ \ P^A_{\perp B}e_A^{\ \mu}=0 \ .
\eea

Given any bulk tensor, e.g. $T^{AB\ldots}_{\ \ C\ldots}$, we can make a normal tensor by projecting it,
\be 
T^{\perp AB\ldots}_{\ \ \ \ C\ldots}=P^A_{\perp D}P^B_{\perp E}P^F_{\perp C}\cdots T^{DE\ldots}_{\ \  F\ldots} \ .
\ee
A tensor is normal if and only if it is equal to its normal projection.  

We have the relations 
\be 
n^A_{\ i}n^{\ i}_{B}=P^A_{\perp B} \ .
\ee
\be 
P^A_{\perp C}P^C_{\ B}=P^A_{\ C}P^C_{\perp B}=0 \ .
\ee
\be 
P^A_{\perp B}+P^A_{\ B}=\delta^A_{\ B}\ .
\ee

We may also define mixed tensors, with some indices tangent and others normal.  Such a tensor $T^{A\cdots\ \ C\cdots}_{\ \ \ B\cdots\  \ D\cdots}$, where the first group of indices $A\cdots$, $B\cdots$ are to be tangent and the second group $C\cdots$, $D\cdots$ are to be normal, is one that admits the decomposition
\be 
T^{A\cdots\ \ C\cdots}_{\ \ \ B\cdots\  \ D\cdots}=T^{\mu\cdots\ \ i\cdots}_{\ \ \ \nu\cdots\  \ j\cdots}e^A_{\ \mu}\cdots e_B^{\ \nu}\cdots n^C_{\ i}\cdots n_D^{\ j}\cdots \ 
\ee

A general tensor can always be decomposed into parallel, normal, and mixed components.  For example, a general $(1,1)$ tensor $T^A_{\ B}$ can be written
\be 
T^A_{\ B}=T^\mu_{\ \nu}e^A_{\ \mu}e_B^{\ \nu}+T^\mu_{\ i}e^A_{\ \mu}n_B^{\ i}+T^i_{\ \mu}n^A_{\ i}e_B^{\ \mu}+T^i_{\ j}n^A_{\ i}n_B^{\ j} \ .
\ee

\subsection{Induced connections}

Consider now the covariant derivatives of a vector in the parallel directions.  This is a quantity which is well defined on the brane itself, i.e. the vector need only be defined on the brane.  Starting from the covariant derivatives of a parallel vector in the parallel directions, we may expand the result into tangent and normal directions via the Gauss-Weingarten relation
\be
{ e^B_{\ \mu}\nabla_B e^A_{\ \nu}=\Gamma^\rho_{\mu\nu}e^A_{\ \rho}-K^i_{\mu\nu}n^A_{\ i}} \ . 
\ee
Here $\Gamma^\rho_{\mu\nu}$ and $K^i_{\mu\nu}$ are defined as the expansion coefficients, equal to 
\be 
\Gamma^\rho_{\mu\nu}=e^{\ \rho}_Ae^B_{\ \mu}\nabla_B e^A_{\ \nu} \ , 
\ee
\be 
{ K^i_{\mu\nu}=-n_A^{\ i} e^B_{\ \mu}\nabla_B e^A_{\ \nu}} \ .
\ee

It is straightforward to show that $\Gamma^\rho_{\mu\nu}$ transforms as a connection under changes in the brane coordinates $x^\mu$, and it is in fact precisely the Levi-Civita connection of the induced metric $g_{\mu\nu}$,
\be 
\Gamma^\rho_{\mu\nu}={1\over 2}g^{\rho\lambda}\left(\partial_\mu g_{\nu\lambda}+\partial_\nu g_{\lambda\mu}-\partial_\lambda g_{\mu\nu}\right) \ .
\ee

The quantity $K^i_{\mu\nu}$ transforms as a tensor in its $\mu\nu$ indices under changes in the brane coordinates, and as a vector in its $i$ index under orthogonal changes in the frame $n^A_{\ i}$.  It is called the \textit{extrinsic curvature}.  By using the relation $\nabla_{A}(n_B^{\ i} e^B_{\ \nu})=0$, we can also write it as 
\be
{ K^i_{\mu\nu}\equiv \nabla_B n_A^{\ i} \ e^B_{\ \mu} e^A_{\ \nu}} \ .
\ee
The extrinsic curvature is symmetric
\be 
K^i_{\mu\nu}=K^i_{\nu\mu} \ ,
\ee
which can be easily shown by noting that the basis vectors have zero lie bracket, hence $e^B_{\ \nu}\nabla_B e^A_{\ \mu}=e^B_{\ \mu}\nabla_B e^A_{\ \nu}.$  We also have
\be 
K^i_{\mu\nu}=\nabla_{(A}n_{B)}^{\ i}e^A_{\ \mu}e^B_{\ \nu}=\half e^A_{\ \mu}e^B_{\ \nu} \mathcal{L}_{n_i}G_{AB} \ .
\ee
Its trace is given by 
\be 
K^i=g^{\mu\nu}K^i_{\mu\nu}=\nabla_A n^{A i} \ .
\ee
Note that in higher co-dimension, the extrinsic curvature gains another index, $i$.  There is one extrinsic curvature component for each normal direction.

Next consider the covariant derivatives of a normal vector in the parallel directions, and expand the result into normal and tangent directions
\be
{ e^B_{\ \mu}\nabla_B n^A_{\ i}=\beta^j_{\mu i}n^A_{\ j}+K_{i\mu}^{\ \ \nu}e^{A}_{\ \nu}} \ .
\ee
Here $\beta^j_{\mu i}$ and $K_{i\mu}^{\ \ \nu}$ are defined as the expansion coefficients, equal to 
\be 
\beta^j_{\mu i}=n_A^{\ j} e^B_{\ \mu}\nabla_B n^A_{\ i} \ , 
\ee
\be 
K_{i\mu}^{\ \ \nu}=e_{A}^{\ \nu} e^B_{\ \mu}\nabla_B n^A_{\ i} \ .
\ee
The $K_{i\mu}^{\ \ \nu}$ are again the extrinsic curvature, with indices raised and lowered as shown.  

The $\beta^j_{\mu i}$ transform as a connection under orthogonal changes in the frame $n^A_{\ i}$.  It is called the \textit{twist connection}, and is the metric connection on the normal bundle, metric compatibility being expressed as the anti-symmetry relation
\be \beta^k_{\mu j}\eta_{ki}=- \beta^k_{\mu i}\eta_{kj}.\ee
The twist connection vanishes identically in co-dimension one, so it is an essentially higher co-dimension object.  

Using the connection on the tangent bundle $\Gamma^\rho_{\mu\nu}$, and the connection $\beta^j_{\mu i}$ on the normal bundle, we can define covariant derivatives $D_\mu$.  Acting on a general mixed tensor $T^{\mu\cdots\ \ i\cdots}_{\ \ \ \nu\cdots\  \ j\cdots}$,
\bea 
D_\rho T^{\mu\cdots\ \ i\cdots}_{\ \ \ \nu\cdots\  \ j\cdots} &=&\partial_\rho T^{\mu\cdots\ \ i\cdots}_{\ \ \ \nu\cdots\  \ j\cdots}+ \Gamma_{\rho\sigma}^\mu T^{\sigma \cdots\ \ i\cdots}_{\ \ \ \nu\cdots\  \ j\cdots} +\cdots \nn \\
&&-\Gamma^\sigma_{\rho\nu} T^{\mu\cdots\ \ i\cdots}_{\ \ \ \sigma\cdots\  \ j\cdots} -\cdots \nn \\
&&+ \beta_{\rho k}^iT^{\mu \cdots\ \ k\cdots}_{\ \ \ \nu\cdots\  \ j\cdots} +\cdots \nn \\
&&-\beta^k_{\rho j} T^{\mu\cdots\ \ i\cdots}_{\ \ \ \nu\cdots\  \ k\cdots} -\cdots \ .
\eea
The covariant derivative $D_\rho T^{\mu\cdots\ \ i\cdots}_{\ \ \ \nu\cdots\  \ j\cdots}$ transforms as a tensor, in the manner indicated by its indices.  

\subsection{Curvatures}

By commutating the covariant derivatives, we arrive at curvature tensors 
\bea
\left[D_\mu,D_\nu\right]T^{\rho\cdots\ \ i\cdots}_{\ \ \ \sigma\cdots\  \ j\cdots}&=& ^{(d)}R^\rho_{\ \lambda \mu\nu}  T^{\lambda\cdots\ \ i\cdots}_{\ \ \ \sigma\cdots\  \ j\cdots}+\cdots \nn \\
&&- ^{(d)}R^\lambda_{\ \sigma \mu\nu}  T^{\rho\cdots\ \ i\cdots}_{\ \ \ \lambda\cdots\  \ j\cdots}-\cdots \nn \\
&&+ ^{(\perp)}R^i_{\ k \mu\nu}  T^{\rho\cdots\ \ k\cdots}_{\ \ \ \sigma\cdots\  \ j\cdots}+\cdots \nn \\
&&-^{(\perp)}R^k_{\ j \mu\nu}T^{\rho\cdots\ \ i\cdots}_{\ \ \ \sigma\cdots\  \ k\cdots}-\cdots \ ,
\eea
 where the curvatures are defined as
\bea 
^{(d)} R^{\rho}_{\ \sigma\mu\nu}&=&\partial_\mu\Gamma^ \rho_{\nu \sigma}-\partial_\nu\Gamma^ \rho_{\mu \sigma}+\Gamma^ \rho_{\mu\lambda}\Gamma^\lambda_{\nu \sigma}-\Gamma^ \rho_{\nu\lambda}\Gamma^\lambda_{\mu \sigma} \ , \\ 
^{(\perp)}R^i_{\ j \mu\nu}&=&\partial_\mu \beta^i_{\nu j}-\partial_\nu\beta^i_{\mu j}+\beta^i_{\mu k}\beta^k_{\nu j}-\beta^i_{\nu k}\beta^k_{\mu j} \ .
\eea
These are anti-symmetric in their first two indices and in their last two indices, and transform as tensors.
 
The bulk curvature components, which can be determined from data localized solely on the brane, can be written in terms of brane quantities.  The relations are the Gauss, Codazzi, and Ricci equations respectively,
 \bea 
 R_{ABCD}e^C_{\ \mu}e^D_{\ \nu}e^A_{\ \rho}e^B_{\ \sigma}&=& 
^{(d)}R_{\rho\sigma\mu\nu}+K^i_{\mu \sigma}K_{i\nu \rho}-K^i_{\nu \sigma}K_{i \mu \rho} \\ 
R_{ABCD}e^C_{\ \mu}e^D_{\ \nu}e^B_{\ \rho}n^{A i}&=& 
D_\nu K^i_{\mu\rho}-D_\mu K^i_{\nu\rho} \\
R_{ABCD}e^C_{\ \mu}e^D_{\ \nu}n^A_{\ j}n^B_{\ i}&=&^{(\perp)}R_{ji \mu\nu}+K_{i\mu}^{\ \ \rho}K_{j\nu\rho}-K_{i\nu}^{\ \ \rho}K_{j\mu\rho} \ .
\eea
 
The final equation only appears in co-dimension $>1$.  Recall that in these expressions the covariant derivative must also act on $i,j\cdots$ indices, via the connection $\beta^i_{\mu j}$.

\section{\label{lovelockappendix}Lovelock terms}

Let the dimension be $D$.  For even $N\geq 2$, define, 
\be {\cal L}^{(N)}={1\over 2^{N/2}}N!\delta^{\mu_1\mu_2\ldots\mu_{N-1}\mu_N}_{\nu_1\nu_2\ldots\nu_{N-1}\nu_N}R_{\mu_1\mu_2}^{\ \ \   \ \ \nu_1\nu_2}R_{\mu_3\mu_4}^{\ \ \ \ \ \nu_3\nu_4}\cdots R_{\mu_{N-1}\mu_N}^{\ \ \ \ \ \ \ \nu_{N-1}\nu_N}.\ee
The delta symbol is defined as 
\be \delta^{\mu_1\mu_2\ldots\mu_{n-1}\mu_n}_{\nu_1\nu_2\ldots\nu_{n-1}\nu_n}\equiv \delta^{[\mu_1}_{\nu_1} \delta^{\mu_2}_{\nu_2}\cdots \delta^{\mu_{n-1}}_{\nu_{n-1}}\delta^{\mu_n]}_{\nu_n}={1\over n!}\left|\begin{array}{ccc}\delta^{\mu_1}_{\nu_1} & \cdots & \delta^{\mu_1}_{\nu_n} \\\vdots & \ddots & \vdots \\\delta^{\mu_n}_{\nu_1} & \cdots & \delta^{\mu_n}_{\nu_n}\end{array}\right|,\ee
It is anti-symmetric in the $\mu$'s, anti-symmetric in the $\nu$'s, and symmetric under the interchange of any $\mu,\nu$ pair with another.  For $n\geq m$ it satisfies the identity
\be \delta^{\mu_1\ldots\mu_n}_{\nu_1\ldots\nu_n}\delta_{\mu_1\ldots\mu_m}^{\nu_1\ldots\nu_m}={(n-m)!\over n!}\left[\prod_{i=1}^m\left(D-(n-i)\right)\right]\delta^{\mu_{m+1}\ldots\mu_n}_{\nu_{m+1}\ldots\nu_n},\ee
as well as identities obtained by expanding out the determinant above in minors, such as the following 
\bea\nn  \delta^{\mu_1\ldots\mu_n}_{\nu_1\ldots\nu_n}&=&{1\over n}\left(\delta^{\mu_1}_{\nu_1}\delta^{\mu_2\ldots\mu_n}_{\nu_2\ldots\nu_n}-\delta^{\mu_1}_{\nu_2}\delta^{\mu_2\ldots\mu_n}_{\nu_1\nu_3\ldots\nu_n}+\cdots+(-1)^n\delta^{\mu_1}_{\nu_n}\delta^{\mu_2\ldots\mu_n}_{\nu_1\ldots\nu_{n-1}}\right)\\ &=&{1\over n}\left(\delta^{\mu_1}_{\nu_1}\delta^{\mu_2\ldots\mu_n}_{\nu_2\ldots\nu_n}-\delta^{\mu_2}_{\nu_1}\delta^{\mu_1\mu_3\ldots\mu_n}_{\nu_2\ldots\nu_n}+\cdots+(-1)^n\delta^{\mu_n}_{\nu_1}\delta^{\mu_1\ldots\mu_{n-1}}_{\nu_2\ldots\nu_n}\right).  \eea

The term ${\cal L}^{(N)}$ vanishes identically for $N<D$ (with $D$ even or odd).  For $D$ even, the integral over a compact oriented riemannian manifold gives the Euler characteristic
\be \chi(M)={1\over (4\pi)^{D/2}\left(D\over 2\right)!}\int d^{D}x\ \sqrt{|g|}\mathcal{L}^{(D)}.\ee 
In particular, this integral does not depend on the metric.  Therefore, for any background metric its variation with respect to the metric must vanish, and thus the integrand must be a total derivative, $\sqrt{|g|}  {\cal L}^{(D)}=\partial_\mu (\rm something)^\mu$.  

The first few terms are 
  \bea \mathcal{L}^{(0)}&=&1,\\ \nn
\mathcal{L}^{(2)}&=&R, \\ \nn
\mathcal{L}^{(4)} &=& R^2 - 4 R_{\mu\nu} R^{\mu\nu}
+ R_{\mu\nu\lambda\sigma} R^{\mu\nu\lambda\sigma}.
\eea
  
\bibliographystyle{apsrev}
\bibliography{multigalileon}

\begin{thebibliography}{29}
\expandafter\ifx\csname natexlab\endcsname\relax\def\natexlab#1{#1}\fi
\expandafter\ifx\csname bibnamefont\endcsname\relax
  \def\bibnamefont#1{#1}\fi
\expandafter\ifx\csname bibfnamefont\endcsname\relax
  \def\bibfnamefont#1{#1}\fi
\expandafter\ifx\csname citenamefont\endcsname\relax
  \def\citenamefont#1{#1}\fi
\expandafter\ifx\csname url\endcsname\relax
  \def\url#1{\texttt{#1}}\fi
\expandafter\ifx\csname urlprefix\endcsname\relax\def\urlprefix{URL }\fi
\providecommand{\bibinfo}[2]{#2}
\providecommand{\eprint}[2][]{\url{#2}}

\bibitem[{\citenamefont{Dvali et~al.}(2000)\citenamefont{Dvali, Gabadadze, and
  Porrati}}]{Dvali:2000hr}
\bibinfo{author}{\bibfnamefont{G.~R.} \bibnamefont{Dvali}},
  \bibinfo{author}{\bibfnamefont{G.}~\bibnamefont{Gabadadze}},
  \bibnamefont{and} \bibinfo{author}{\bibfnamefont{M.}~\bibnamefont{Porrati}},
  \bibinfo{journal}{Phys. Lett.} \textbf{\bibinfo{volume}{B485}},
  \bibinfo{pages}{208} (\bibinfo{year}{2000}), \eprint{hep-th/0005016}.

\bibitem[{\citenamefont{Luty et~al.}(2003)\citenamefont{Luty, Porrati, and
  Rattazzi}}]{Luty:2003vm}
\bibinfo{author}{\bibfnamefont{M.~A.} \bibnamefont{Luty}},
  \bibinfo{author}{\bibfnamefont{M.}~\bibnamefont{Porrati}}, \bibnamefont{and}
  \bibinfo{author}{\bibfnamefont{R.}~\bibnamefont{Rattazzi}},
  \bibinfo{journal}{JHEP} \textbf{\bibinfo{volume}{09}}, \bibinfo{pages}{029}
  (\bibinfo{year}{2003}), \eprint{hep-th/0303116}.

\bibitem[{\citenamefont{Nicolis and Rattazzi}(2004)}]{Nicolis:2004qq}
\bibinfo{author}{\bibfnamefont{A.}~\bibnamefont{Nicolis}} \bibnamefont{and}
  \bibinfo{author}{\bibfnamefont{R.}~\bibnamefont{Rattazzi}},
  \bibinfo{journal}{JHEP} \textbf{\bibinfo{volume}{06}}, \bibinfo{pages}{059}
  (\bibinfo{year}{2004}), \eprint{hep-th/0404159}.

\bibitem[{\citenamefont{Gabadadze and Iglesias}(2006)}]{Gabadadze:2006tf}
\bibinfo{author}{\bibfnamefont{G.}~\bibnamefont{Gabadadze}} \bibnamefont{and}
  \bibinfo{author}{\bibfnamefont{A.}~\bibnamefont{Iglesias}},
  \bibinfo{journal}{Phys. Lett.} \textbf{\bibinfo{volume}{B639}},
  \bibinfo{pages}{88} (\bibinfo{year}{2006}), \eprint{hep-th/0603199}.

\bibitem[{\citenamefont{Nicolis et~al.}(2009)\citenamefont{Nicolis, Rattazzi,
  and Trincherini}}]{Nicolis:2008in}
\bibinfo{author}{\bibfnamefont{A.}~\bibnamefont{Nicolis}},
  \bibinfo{author}{\bibfnamefont{R.}~\bibnamefont{Rattazzi}}, \bibnamefont{and}
  \bibinfo{author}{\bibfnamefont{E.}~\bibnamefont{Trincherini}},
  \bibinfo{journal}{Phys. Rev.} \textbf{\bibinfo{volume}{D79}},
  \bibinfo{pages}{064036} (\bibinfo{year}{2009}), \eprint{0811.2197}.

\bibitem[{\citenamefont{de~Rham and Gabadadze}(2010)}]{deRham:2010gu}
\bibinfo{author}{\bibfnamefont{C.}~\bibnamefont{de~Rham}} \bibnamefont{and}
  \bibinfo{author}{\bibfnamefont{G.}~\bibnamefont{Gabadadze}}
  (\bibinfo{year}{2010}), \eprint{1006.4367}.

\bibitem[{\citenamefont{Deffayet
  et~al.}(2009{\natexlab{a}})\citenamefont{Deffayet, Esposito-Farese, and
  Vikman}}]{Deffayet:2009wt}
\bibinfo{author}{\bibfnamefont{C.}~\bibnamefont{Deffayet}},
  \bibinfo{author}{\bibfnamefont{G.}~\bibnamefont{Esposito-Farese}},
  \bibnamefont{and} \bibinfo{author}{\bibfnamefont{A.}~\bibnamefont{Vikman}},
  \bibinfo{journal}{Phys. Rev.} \textbf{\bibinfo{volume}{D79}},
  \bibinfo{pages}{084003} (\bibinfo{year}{2009}{\natexlab{a}}),
  \eprint{0901.1314}.

\bibitem[{\citenamefont{Deffayet
  et~al.}(2009{\natexlab{b}})\citenamefont{Deffayet, Deser, and
  Esposito-Farese}}]{Deffayet:2009mn}
\bibinfo{author}{\bibfnamefont{C.}~\bibnamefont{Deffayet}},
  \bibinfo{author}{\bibfnamefont{S.}~\bibnamefont{Deser}}, \bibnamefont{and}
  \bibinfo{author}{\bibfnamefont{G.}~\bibnamefont{Esposito-Farese}},
  \bibinfo{journal}{Phys. Rev.} \textbf{\bibinfo{volume}{D80}},
  \bibinfo{pages}{064015} (\bibinfo{year}{2009}{\natexlab{b}}),
  \eprint{0906.1967}.

\bibitem[{\citenamefont{Silva and Koyama}(2009)}]{Silva:2009km}
\bibinfo{author}{\bibfnamefont{F.~P.} \bibnamefont{Silva}} \bibnamefont{and}
  \bibinfo{author}{\bibfnamefont{K.}~\bibnamefont{Koyama}},
  \bibinfo{journal}{Phys. Rev.} \textbf{\bibinfo{volume}{D80}},
  \bibinfo{pages}{121301} (\bibinfo{year}{2009}), \eprint{0909.4538}.

\bibitem[{\citenamefont{Kobayashi}(2010)}]{Kobayashi:2010wa}
\bibinfo{author}{\bibfnamefont{T.}~\bibnamefont{Kobayashi}},
  \bibinfo{journal}{Phys. Rev.} \textbf{\bibinfo{volume}{D81}},
  \bibinfo{pages}{103533} (\bibinfo{year}{2010}), \eprint{1003.3281}.

\bibitem[{\citenamefont{De~Felice and Tsujikawa}(2010)}]{DeFelice:2010pv}
\bibinfo{author}{\bibfnamefont{A.}~\bibnamefont{De~Felice}} \bibnamefont{and}
  \bibinfo{author}{\bibfnamefont{S.}~\bibnamefont{Tsujikawa}}
  (\bibinfo{year}{2010}), \eprint{1007.2700}.

\bibitem[{\citenamefont{Chow and Khoury}(2009)}]{Chow:2009fm}
\bibinfo{author}{\bibfnamefont{N.}~\bibnamefont{Chow}} \bibnamefont{and}
  \bibinfo{author}{\bibfnamefont{J.}~\bibnamefont{Khoury}},
  \bibinfo{journal}{Phys. Rev.} \textbf{\bibinfo{volume}{D80}},
  \bibinfo{pages}{024037} (\bibinfo{year}{2009}), \eprint{0905.1325}.

\bibitem[{\citenamefont{Agarwal et~al.}(2010)\citenamefont{Agarwal, Bean,
  Khoury, and Trodden}}]{Agarwal:2009gy}
\bibinfo{author}{\bibfnamefont{N.}~\bibnamefont{Agarwal}},
  \bibinfo{author}{\bibfnamefont{R.}~\bibnamefont{Bean}},
  \bibinfo{author}{\bibfnamefont{J.}~\bibnamefont{Khoury}}, \bibnamefont{and}
  \bibinfo{author}{\bibfnamefont{M.}~\bibnamefont{Trodden}},
  \bibinfo{journal}{Phys. Rev.} \textbf{\bibinfo{volume}{D81}},
  \bibinfo{pages}{084020} (\bibinfo{year}{2010}), \eprint{0912.3798}.

\bibitem[{\citenamefont{Creminelli et~al.}(2010)\citenamefont{Creminelli,
  Nicolis, and Trincherini}}]{Creminelli:2010ba}
\bibinfo{author}{\bibfnamefont{P.}~\bibnamefont{Creminelli}},
  \bibinfo{author}{\bibfnamefont{A.}~\bibnamefont{Nicolis}}, \bibnamefont{and}
  \bibinfo{author}{\bibfnamefont{E.}~\bibnamefont{Trincherini}}
  (\bibinfo{year}{2010}), \eprint{1007.0027}.

\bibitem[{\citenamefont{Gannouji and Sami}(2010)}]{Gannouji:2010au}
\bibinfo{author}{\bibfnamefont{R.}~\bibnamefont{Gannouji}} \bibnamefont{and}
  \bibinfo{author}{\bibfnamefont{M.}~\bibnamefont{Sami}},
  \bibinfo{journal}{Phys. Rev.} \textbf{\bibinfo{volume}{D82}},
  \bibinfo{pages}{024011} (\bibinfo{year}{2010}), \eprint{1004.2808}.

\bibitem[{\citenamefont{Deffayet
  et~al.}(2010{\natexlab{a}})\citenamefont{Deffayet, Pujolas, Sawicki, and
  Vikman}}]{Deffayet:2010qz}
\bibinfo{author}{\bibfnamefont{C.}~\bibnamefont{Deffayet}},
  \bibinfo{author}{\bibfnamefont{O.}~\bibnamefont{Pujolas}},
  \bibinfo{author}{\bibfnamefont{I.}~\bibnamefont{Sawicki}}, \bibnamefont{and}
  \bibinfo{author}{\bibfnamefont{A.}~\bibnamefont{Vikman}}
  (\bibinfo{year}{2010}{\natexlab{a}}), \eprint{1008.0048}.

\bibitem[{\citenamefont{de~Rham and Tolley}(2010)}]{deRham:2010eu}
\bibinfo{author}{\bibfnamefont{C.}~\bibnamefont{de~Rham}} \bibnamefont{and}
  \bibinfo{author}{\bibfnamefont{A.~J.} \bibnamefont{Tolley}},
  \bibinfo{journal}{JCAP} \textbf{\bibinfo{volume}{1005}}, \bibinfo{pages}{015}
  (\bibinfo{year}{2010}), \eprint{1003.5917}.

\bibitem[{\citenamefont{Deffayet
  et~al.}(2010{\natexlab{b}})\citenamefont{Deffayet, Deser, and
  Esposito-Farese}}]{Deffayet:2010zh}
\bibinfo{author}{\bibfnamefont{C.}~\bibnamefont{Deffayet}},
  \bibinfo{author}{\bibfnamefont{S.}~\bibnamefont{Deser}}, \bibnamefont{and}
  \bibinfo{author}{\bibfnamefont{G.}~\bibnamefont{Esposito-Farese}}
  (\bibinfo{year}{2010}{\natexlab{b}}), \eprint{1007.5278}.

\bibitem[{\citenamefont{Padilla
  et~al.}(2010{\natexlab{a}})\citenamefont{Padilla, Saffin, and
  Zhou}}]{Padilla:2010ir}
\bibinfo{author}{\bibfnamefont{A.}~\bibnamefont{Padilla}},
  \bibinfo{author}{\bibfnamefont{P.~M.} \bibnamefont{Saffin}},
  \bibnamefont{and} \bibinfo{author}{\bibfnamefont{S.-Y.} \bibnamefont{Zhou}}
  (\bibinfo{year}{2010}{\natexlab{a}}), \eprint{1008.0745}.

\bibitem[{\citenamefont{Padilla
  et~al.}(2010{\natexlab{b}})\citenamefont{Padilla, Saffin, and
  Zhou}}]{Padilla:2010de}
\bibinfo{author}{\bibfnamefont{A.}~\bibnamefont{Padilla}},
  \bibinfo{author}{\bibfnamefont{P.~M.} \bibnamefont{Saffin}},
  \bibnamefont{and} \bibinfo{author}{\bibfnamefont{S.-Y.} \bibnamefont{Zhou}}
  (\bibinfo{year}{2010}{\natexlab{b}}), \eprint{1007.5424}.

\bibitem[{\citenamefont{Cadoni and Pani}(2009)}]{Cadoni:2008fb}
\bibinfo{author}{\bibfnamefont{M.}~\bibnamefont{Cadoni}} \bibnamefont{and}
  \bibinfo{author}{\bibfnamefont{P.}~\bibnamefont{Pani}},
  \bibinfo{journal}{Phys. Lett.} \textbf{\bibinfo{volume}{B674}},
  \bibinfo{pages}{308} (\bibinfo{year}{2009}), \eprint{0812.3010}.

\bibitem[{\citenamefont{Lovelock}(1971)}]{Lovelock:1971yv}
\bibinfo{author}{\bibfnamefont{D.}~\bibnamefont{Lovelock}},
  \bibinfo{journal}{J. Math. Phys.} \textbf{\bibinfo{volume}{12}},
  \bibinfo{pages}{498} (\bibinfo{year}{1971}).

\bibitem[{\citenamefont{Dyer and Hinterbichler}(2009)}]{Dyer:2008hb}
\bibinfo{author}{\bibfnamefont{E.}~\bibnamefont{Dyer}} \bibnamefont{and}
  \bibinfo{author}{\bibfnamefont{K.}~\bibnamefont{Hinterbichler}},
  \bibinfo{journal}{Phys. Rev.} \textbf{\bibinfo{volume}{D79}},
  \bibinfo{pages}{024028} (\bibinfo{year}{2009}), \eprint{0809.4033}.

\bibitem[{\citenamefont{Gibbons and Hawking}(1977)}]{Gibbons:1976ue}
\bibinfo{author}{\bibfnamefont{G.~W.} \bibnamefont{Gibbons}} \bibnamefont{and}
  \bibinfo{author}{\bibfnamefont{S.~W.} \bibnamefont{Hawking}},
  \bibinfo{journal}{Phys. Rev.} \textbf{\bibinfo{volume}{D15}},
  \bibinfo{pages}{2752} (\bibinfo{year}{1977}).

\bibitem[{\citenamefont{York}(1972)}]{York:1972sj}
\bibinfo{author}{\bibfnamefont{J.~W.} \bibnamefont{York}, \bibfnamefont{Jr.}},
  \bibinfo{journal}{Phys. Rev. Lett.} \textbf{\bibinfo{volume}{28}},
  \bibinfo{pages}{1082} (\bibinfo{year}{1972}).

\bibitem[{\citenamefont{Myers}(1987)}]{Myers:1987yn}
\bibinfo{author}{\bibfnamefont{R.~C.} \bibnamefont{Myers}},
  \bibinfo{journal}{Phys. Rev.} \textbf{\bibinfo{volume}{D36}},
  \bibinfo{pages}{392} (\bibinfo{year}{1987}).

\bibitem[{\citenamefont{Miskovic and Olea}(2007)}]{Miskovic:2007mg}
\bibinfo{author}{\bibfnamefont{O.}~\bibnamefont{Miskovic}} \bibnamefont{and}
  \bibinfo{author}{\bibfnamefont{R.}~\bibnamefont{Olea}},
  \bibinfo{journal}{JHEP} \textbf{\bibinfo{volume}{10}}, \bibinfo{pages}{028}
  (\bibinfo{year}{2007}), \eprint{0706.4460}.

\bibitem[{\citenamefont{Charmousis and
  Zegers}(2005{\natexlab{a}})}]{Charmousis:2005ey}
\bibinfo{author}{\bibfnamefont{C.}~\bibnamefont{Charmousis}} \bibnamefont{and}
  \bibinfo{author}{\bibfnamefont{R.}~\bibnamefont{Zegers}},
  \bibinfo{journal}{JHEP} \textbf{\bibinfo{volume}{08}}, \bibinfo{pages}{075}
  (\bibinfo{year}{2005}{\natexlab{a}}), \eprint{hep-th/0502170}.

\bibitem[{\citenamefont{Charmousis and
  Zegers}(2005{\natexlab{b}})}]{Charmousis:2005ez}
\bibinfo{author}{\bibfnamefont{C.}~\bibnamefont{Charmousis}} \bibnamefont{and}
  \bibinfo{author}{\bibfnamefont{R.}~\bibnamefont{Zegers}},
  \bibinfo{journal}{Phys. Rev.} \textbf{\bibinfo{volume}{D72}},
  \bibinfo{pages}{064005} (\bibinfo{year}{2005}{\natexlab{b}}),
  \eprint{hep-th/0502171}.

\end{thebibliography}

\end{document}